# Comparing NEO Search Telescopes

Nathan Myhrvold

Jan 19th, 2016

## Abstract

Multiple terrestrial and space-based telescopes have been proposed for detecting and tracking near-Earth objects (NEOs). Detailed simulations of the search performance of these systems have used complex computer codes that are not widely available, which hinders accurate cross-comparison of the proposals and obscures whether they have consistent assumptions. Moreover, some proposed instruments would survey infrared (IR) bands, whereas others would operate in the visible band, and differences among asteroid thermal and visible light models used in the simulations further complicate like-to-like comparisons. I use simple physical principles to estimate basic performance metrics for the ground-based Large Synoptic Survey Telescope and three space-based instruments—Sentinel, NEOCam, and a Cubesat constellation. The performance is measured against two different NEO distributions, the Bottke *et al.* distribution of general NEOs, and the Veres *et al.* distribution of Earth impacting NEO. The results of the comparison show simplified relative performance metrics, including the expected number of NEOs visible in the search volumes and the initial detection rates expected for each system. Although these simplified comparisons do not capture all of the details, they give considerable insight into the physical factors limiting performance. Multiple asteroid thermal models are considered, including FRM, NEATM, and a new generalized form of FRM (GFRM). I describe issues with how IR albedo and emissivity have been estimated in previous studies, which may render them inaccurate. A thermal model for tumbling asteroids is also developed and suggests that tumbling asteroids may be surprisingly difficult for IR telescopes to observe.

## 1. Introduction

Near Earth objects have intrinsic scientific interest as well as practical significance because large asteroid impacts on Earth could cause widespread destruction. The NASA Authorization Act of 2005 gave NASA the objective of detecting, tracking, cataloguing, and characterizing 90% of NEOs of diameter $d \geq 140$ m and perihelion $ph < 1.3$ by 2020, known as the George E. Brown Near Earth Object Survey (hereinafter, GEB). This objective did not specify a technical approach nor did it allocate funding. Since then, several projects have been proposed to address this need. But delays in funding and approval make it unlikely that any project can meet the 2020 deadline.

Construction has begun on the Large Synoptic Survey Telescope (LSST) (Ivezić et al., 2008), which has a wide field of view and an 8.4 m primary mirror (6.4 m net of secondary mirror obstruction). The LSST was designed to make a deep survey of the entire sky, and while doing that it would also find NEOs as faint as apparent Johnson visual magnitude 24.6. Simulations by the LSST team (Ivezić et al., 2006) show that the LSST should be able to complete the GEB in about 12 years of observation—potentially faster if a NEO-optimized observing cadence is adopted.

The Sentinel Mission is sponsored by the B612 Foundation and Ball Aerospace as a space-based telescope operating in the infrared with a 0.5 meter primary mirror, on a Venus-like orbit (Arentz et al., 2010; Buie and Reitsema, 2015; Lu et al., 2013). The B612 Foundation has proposed raising private donations to fund Sentinel.

NEOCam is an infrared space telescope of similar size proposed by the Jet Propulsion Laboratory as a NASA-funded Discovery mission (Mainzer et al., 2015). NEOCam has a different pass band and observing strategy than Sentinel, and it would orbit at the L1 Lagrange point of the Sun-Earth system.

In 2010, the National Research Council published a report that reviewed the designs of NEO-detection systems that have been fielded as well as the proposals for LSST, Sentinel, and NEOCam (Shapiro et al., 2010). The NRC concluded that LSST offered the most cost-effective and lowest risk approach, but that it would not complete the GEB by 2020. Space-based approaches were judged to be more expensive and riskier, but if cost were no object, they could supplement LSST to complete the GEB more quickly.

In the time since the National Academy report, each project has continued to refine its approach. LSST received its final funding on the basis of non-NEO scientific goals, and construction started in 2015. Under its current schedule, LSST will achieve first light in 2020 and will begin observing in 2022. LSST funding is currently sufficient for 10 years of observations, which is enough time to make substantial progress on the GEB, but not to complete it. The LSST team has proposed additional funding to extend its mission to 12 years and adopt a NEO-optimized observing cadence in order to complete the GEB. However even with its current funding, the LSST will find many NEOs. The marginal cost of performing NEO searches using LSST is very low.

A fourth proposal, released after the National Academy report, is a space-based constellation of five small (~9 liter volume) Cubesats (Shao et al., 2015). This innovative approach has a much lower projected cost ($50 m, about 10% of the cost of Sentinel or NEOCam). Failure of any one satellite would reduce search performance but not lead to mission failure. These are very attractive features if borne out in practice. However, given that LSST is already funded, it is unclear what the role there is for any space-based NEO mission unless it exceeds LSST in some part of the performance-parameter space.

In addition to the four telescopes studied here, many other current or proposed systems have been described. Space telescopes include WISE/NEOWISE (Mainzer et al., 2014a, 2011b; Wright et al., 2010) and NEOSsat (Greenstreet and Gladman, 2013; Greenstreet et al., 2012). Among the ground-based surveys, Pan-STARRS (Denneau et al., 2013), the Catalina Sky Survey (Vereš et al., 2009), and the Space Surveillance Telescope survey (Ruprecht et al., 2014) are operating, ATLAS (Tonry, 2011) is under construction, and ADAM (Vereš et al., 2014) is an active proposal. These systems are not covered here because they either represent the previous generation of NEO surveys (in the case of NEOWISE, Pan-STARRS, and Catalina) or are aimed at different problems, such as late-stage warning of imminent Earth impact (in the case of ATLAS and ADAM). NEOSSat is a recently deployed Canadian space telescope, which is similar in some aspects to the Cubesats, but lacks synthetic tracking. Because NEOSSat has such a narrow field of regard, it is not capable of conducting a comprehensive NEO survey.

The GEB objective is concerned solely with objects of at least 140m in diameter. The NRC concluded, however, that "NEOs as small as 30 to 50 meters in diameter could be highly destructive," and it urged that searches thus also attempt to track Earth-impacting asteroids smaller than 140 m (Shapiro et al., 2010). In the 1908 Tunguska event, an object of estimated 20 to 50 m diameter caused devastation for hundreds of square miles(Boslough and Crawford, 2008; Chyba et al., 1993)More recently, the 2013 Chelyabinsk bolide—estimated at just 19 m in diameter (Brown et al., 2013; Kring et al., 2013)—broke >1 million windows and caused more than 1000 injuries (Popova et al., 2013). Damage would likely have been even more severe had the impact angle not been so shallow (estimated at 18 degrees above the horizon). As I show here, some of the current proposals would enable detection of some asteroids smaller than 140 m.

The current best practice for assessing the performance of each instrument is to perform Monte Carlo simulations that create a hypothetical test population of NEO orbits by sampling orbits from either de-biased general populations of NEOs (Bottke et al., 2002) or from a subset of NEOs whose orbits have been computed to impact Earth (Chesley and Spahr, 2004; Vereš et al., 2009). Such simulations have been performed for LSST (Ivezić et al., 2006) and several of the other proposed systems (Arentz et al., 2010; Buie and Reitsema, 2015; Mainzer et al., 2015; Shao et al., 2015). These models can take into account any details that their creators feel are important, including weather-related observing outages in the case of ground-based telescopes (Ivezić et al., 2006).

Such simulations are extremely important, but they also have some shortcomings. To date, simulations have typically been implemented by the teams responsible for each proposed mission, and the codes are not available to other researchers. A research team that has a model of its instrument can add features to simulate other, competing instruments (Buie and Reitsema, 2015; Grav et al., 2015)but those added features may not capture all of the important details of the alternative system. Research groups not connected to projects have also performed important simulations (Moon et al., 2008; Vereš et al., 2009), but no such study covers the four systems studied here.

In addition to performance comparison with other projects, simulations also perform an important role in shaping the design tradeoffs of a program because the simulation is used to determine performance of different options internal to a project. To a large extent the key design decisions are made on the basis of simulation output, reinforcing the need for simulations that are both detailed and accurate.

The more detailed a simulation is, the better it is able to model the actual performance of the system, but very complicated models can be opaque to intuition. Published results of simulations typically include only a limited range of metrics, such as the time to complete the GEB, and often do not include performance metrics such as detection capabilities for asteroids smaller than 140 m, nor do they always list underlying assumptions comprehensively. As a result, it is difficult for independent groups to compare the relative performance of proposed NEO detection systems in all regimes of interest.

Ideally, the community would produce an open model that can simulate the NEO search performance of IR and visible-light telescopes, whether based on the ground or in space, with consistent assumptions and consistent input distributions of NEOs.

Absent such a model, it is useful to have simplified metrics that allow one to compare the most important features of each NEO telescope. I calculate such metrics here. Because these calculations are unable to account for many important details, these results should be considered crude approximations rather than definitive explorations of the telescopes' performance. Nevertheless, I show here that this approach does yield broad-brush comparisons that highlight important issues, which more detailed simulations can focus on.

When comparing telescopes, one must reconcile the differing methods of analysis and baseline assumptions used by each team, including the basics of optical telescope visibility and asteroid thermal models. While performing this analysis, I found that some previous analyses used inconsistent assumptions or incorrect models.

## 2. Methods

### 2.1. Asteroid Distributions

Bottke et al., 2002 created a de-biased statistical distribution of orbital elements eccentricity $e$, semimajor axis $a$, and inclination $i$ for NEOs. This distribution, which has been widely used for evaluating NEO searches, assumes that the longitude of the ascending node $\Omega$ and the argument of periapsis $\omega$ are uniformly distributed, and that the orbits are independent of asteroid size. Although newer estimates of the NEO population have recently become available(Harris and D'Abramo, 2015), the Bottke distribution is used here because it is the basis of the simulations used in each of the telescope specific studies (Arentz et al., 2010; Buie and Reitsema, 2015; Ivezić et al., 2006; Mainzer et al., 2015; Shao et al., 2015).

As published, the Bottke *et al.* distribution gives probabilities for a coarse grid in $e, a, i$. I obtained a general NEO distribution by linearly interpolating the probability density and sub-sampling on a grid over a finer-resolution $e, a, i$ grid constrained to aphelion $q \geq 0.983$ and perihelion $0 \leq Q \leq 1.3$, both in au. This resulted in 10,518 $(e, a, i)$ triplets having non-zero probability. This distribution is appropriate for the GEB.

Vereš et al., 2009, building on earlier work by Chesley and Spahr, 2004, created a second distribution limited to asteroids that appear to be on a collision course with Earth. To generate this distribution, Veres *et al.* drew asteroids at random from the Bottke *et al.* distribution and then evolved the system forward in time, using an accurate solar-system simulator. Non-impactors were discarded from the distribution. The resulting impactor distribution includes 10,006 orbits (Vereš et al., 2009). The aphelion and perihelion values for both distributions are shown in Fig. 1.

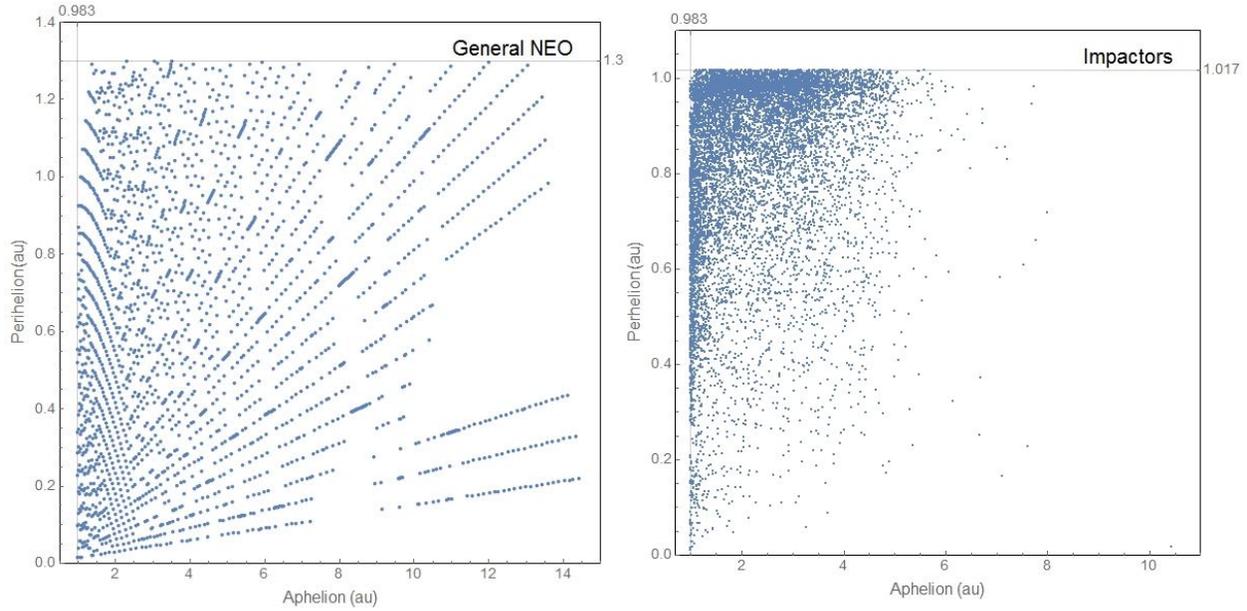

**Fig. 1.** Aphelion and perihelion values for both the general NEO distribution (left) and impactor distribution (right). The general NEO distribution was sampled in eccentricity, semi-major axis, and inclination ($e, a, i$), and then converted to values of perihelion and aphelion. Each point is associated with a probability value. Only 1/5 of the points are shown. The impactor distribution is shown as actual orbits, with one point per orbit.

Given orbital elements describing an asteroid orbit, one can apply Kepler's second law to find the probability distribution for an asteroid at a given radius $r$ from the Sun. Fig. 2 shows an example orbit and the resulting radial distributions of general NEOs and impactors. Note that the radial distributions are independent of elements $i, \Omega, \omega$; they depend only on aphelion and perihelion, or equivalently on $a$ and $e$.

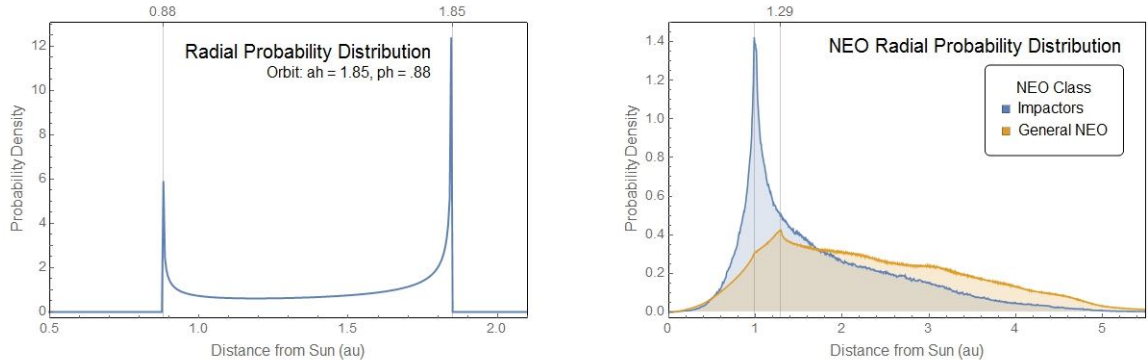

*Fig. 2. Probability of finding a NEO a distance r from the Sun. Left: radial probability density function for a single orbit having ah=1.85 au and ph=0.88 au. Right: summed and normalized distributions of asteroids by radial distance from the Sun across the 10 006 impactor orbits derived by Vereš et al., 2009 and an 11 288-orbit sample of the general NEO distribution published by Bottke et al., 2002.*

The general NEO distribution peaks at $r = 1.29$ au, and 89% of the distribution falls within $r \geq 1$. The impactor distribution peaks at $r = 0.983$ au; 79% of the distribution falls within $r \geq 1$.

Given a sufficiently large number of asteroids, one would expect these distributions to be stationary—as NEOs proceed around their orbits, the distribution should remain nearly unchanged.

## 2.2. Asteroids in Reflected Light

The simplest model of an asteroid in reflected light is a diffuse Lambertian sphere. Under this assumption, an asteroid of diameter $d$ (in meters), geometric albedo $p_v$, and phase angle $\alpha = 0$ (*i.e.*, at opposition) will have visual absolute magnitude $H$ given by

$$H(d, p_v) = 5 \operatorname{Log}_{10}\left(\frac{1{,}329{,}000}{d\sqrt{p_v}}\right). \tag{1}$$

Real asteroids are neither spherical nor perfectly Lambertian, but this formula is nonetheless widely used to estimate $p_v$ for asteroids, given measured values of $H$ and estimates for $d$.

The phase function $\psi(\alpha)$ describes how the light falls off with non-zero phase angle. The phase integral $q$, which is very important for energy balance, is given by

$$q = 2 \int_0^\pi \sin\alpha\, \psi(\alpha)\, d\alpha. \tag{2}$$

I use the HG system, a phase function derived from empirical observations (Bowell et al., 1989), to model the non-Lambertian aspects of the hypothetical asteroids.

$$\psi_{\mathrm{HG}}(\alpha, G) = \\
e^{-90.56 \tan\left(\frac{\alpha}{2}\right)^2} - (G-1)\left(e^{-3.332 \tan\left(\frac{\alpha}{2}\right)^{0.631}} + e^{-3.332 \tan\left(\frac{\alpha}{2}\right)^{0.631} - 90.56 \tan\left(\frac{\alpha}{2}\right)^2}\right) + \\
G\left(e^{-1.862 \tan\left(\frac{\alpha}{2}\right)^{1.218}} - e^{-1.862 \tan\left(\frac{\alpha}{2}\right)^{1.218} - 90.56 \tan\left(\frac{\alpha}{2}\right)^2}\right) + \\
\frac{(1.31 - 0.992\, G) \sin\alpha\, e^{-90.56 \tan\left(\frac{\alpha}{2}\right)^2}}{-0.158 + (-1.79 + \sin\alpha)\sin\alpha}, \tag{3}$$

Where $G$ is an empirically fit constant, which typically is in the range $0.1 \leq G \leq 0.5$. $G$ is roughly correlated to albedo, with lower albedo having lower values of $G$ (Morbidelli et al., 2002). Using a 583-asteroid data set of $G$ and $p_v$ (Pravec et al., 2012), I found a least-squares fit $G = g(p_v) = 0.507\, p_v^{0.477}$, but there is considerable scatter ($R^2 = 0.33$).

A simpler version of the HG phase function is often in use:

$$\psi_{\mathrm{HGs}}(\alpha, G) = (1-G)\, e^{-3.33 \operatorname{Tan}\left(\frac{\alpha}{2}\right)^{0.63}} + G e^{-1.87 \operatorname{Tan}\left(\frac{\alpha}{2}\right)^{1.22}}.$$

The phase integral is widely reported as $q_{\mathrm{HG}}(G) = 0.290 + 0.690\, G$. However when I perform accurate numerical integration of (3), I get

$$q_{\mathrm{HG}}(G) = 0.286 + 0.656\, G, \tag{4}$$

and I used this expression for the analyses presented here.

The Shevchenko phase function (Belskaya and Shevchenko, 2000; Cellino and Gil-Hutton, 2011; Petrova and Tishkovets, 2011) is a purely empirical function that can obtain better fits than the HG and other models do. This function is usually used only for small phase angles $\alpha \leq 30°$. The original Shevchenko function has two parameters, and when expressed in terms of visual magnitude $V$, it is

$$V = V_0 - \frac{a}{1+\alpha} + \alpha\, b,$$

where $a$ controls the opposition effect, and $b$ is the slope. I recast this as a phase function, and replaced $a$ with

$$\rho = \frac{V_{\alpha=0°}}{V_{\alpha=5°}}.$$

Typical values are $1.22 \leq \rho \leq 1.5$ and $0.019 \leq b \leq 0.046$ for asteroids studied to date. Note that from an observational standpoint, $\rho$ is relevant only at angles within a few degrees of opposition, so the most important parameter is $b$. There is evidence(Belskaya and Shevchenko, 2000; Cellino and Gil-Hutton, 2011; Petrova and Tishkovets, 2011) that $b$ has a strong correlation to geometric albedo $p_v$: $b = 0.013 - 0.024\, \mathrm{Log}_{10}(p_v)$. This evidence is based on a fairly small set of asteroids, however.

When recast in the same form as the other phase functions,

$$\psi_{\mathrm{Shevchenko}}(\alpha, \rho, b) = \mathrm{e}^{-0.921\left(\frac{180\,\alpha\,b}{\pi} + \frac{b\,(-0.0314\,-\,0.524\rho)}{(0.0175+\alpha)\,(4.62+\rho)} - \frac{57.3\,b\,(-0.0314\,-\,0.524\rho)}{(4.62+\rho)}\right)}. \tag{5}$$

The HG system is the mostly widely used phase curve in asteroid studies because it was an IAU standard for many decades. The HG system does not produce a good fit to all observed asteroid phase curves, this has led others to create more complicated phase functions, as well as better theoretical models (Caminiti et al., 2014; Jian-Yang Li, 2005; La Forgia F., 2014; Muinonen et al., 2012, 2010a, 2010b)

Unfortunately, theoretically inspired models having many empirically derived parameters are useful for a prospective NEO study only if statistical estimates exist for those parameters across the NEO population.

The HG system has been superseded by the $H, G_1, G_2$ system (Muinonen et al., 2012, 2010a, 2010b), which is the new IAU standard. Measured values for the parameters $G_1, G_2$ only exist for a relatively small fraction of known asteroids (Oszkiewicz et al., 2011), and they have not been used in previous NEO simulation studies.

In this study I use the HG system because it has been used in all of the telescope-specific previous studies of LSST, NEOCam, Sentinel and the Cubesat constellation. We can estimate the visible apparent magnitude at the telescope as

$$V(d, G, p_v, r_{\mathrm{as}}, r_{\mathrm{ao}}, \alpha) = H(d, p_v) + 5\, \mathrm{Log}_{10}(r_{\mathrm{as}}\, r_{\mathrm{ao}}) - 2.5\, \mathrm{Log}_{10}(\psi(\alpha, G)) \tag{6}$$

$$\cos \alpha = \frac{r_{ao}^2 + r_{as}^2 - r_{os}^2}{2\, r_{as}\, r_{ao}} = \vec{s} \cdot \vec{t} \tag{7}$$

$$\sin \gamma = \frac{r_{as} \cos \alpha}{r_{os}},$$

where $V$ is the apparent magnitude, $H$ is the absolute magnitude, and the geometric parameters (see Fig. 4a) are: $r_{ao}$ is the distance between the asteroid and observer in au, $r_{as}$ is the distance between the asteroid and the Sun in au, $r_{os}$ is the distance between the observatory and the Sun in au, $\vec{s}$ is a unit vector from the Sun to the asteroid, $\vec{t}$ is a unit vector from the asteroid to the telescope, and $\gamma$ is the solar elongation angle.

Eq. (6) can be used compute the search volume in which an asteroid of absolute magnitude $H$ can be observed by setting $V = V_{\text{limit}}$, the limiting magnitude detectable at the telescope, and then solving for $r_{as}$ and $r_{ao}$. As Fig. 3 illustrates, the HG system can produce search volumes that span the range of results produced by Lambert and Shevchenko model by simply selecting values of $G$ in the range $0.13 \leq G \leq 0.4$.

The difference between the phase functions in Fig. 3 is not as much as one might imagine for two reasons. First, as shown in Fig. 4b-d, the vast majority of the night-side viewing angles have low values (*i.e.*, $\alpha \leq 60°$). As a result, the effects of phase angle are typically relatively weak.

Second, the inverse-square law tends to diminish the phase effect. When an asteroid is at opposition ($\alpha = 0$), then $\psi(0, G) = 1$, so the phase contribution is zero, but the inverse-square term $5 \log_{10}(r_{as} r_{ao})$ in Eq. (6) is at its maximum because the Sun, asteroid and observer are all in a line (*i.e.*, $r_{as} = r_{ao} + r_{os}$). At $\alpha > 0$, the phase term $-2.5 \log_{10}(\psi(\alpha, G))$ in (6) makes the apparent magnitude dimmer, but the consequently smaller values of $r_{as}$ and $r_{ao}$ partly balance it.

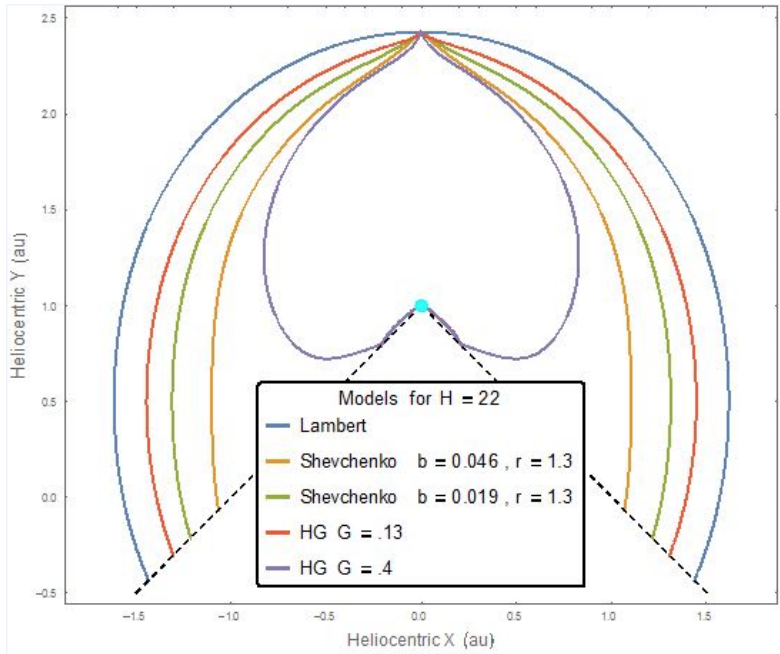

**Fig. 3.** Search volumes computed using Eq. (6) and the Lambert, Shevchenko, and HG phase functions. The observer in this case is at the pale blue dot, and each of the functions shows the boundary in heliocentric $x, y$ coordinates at which an asteroid of absolute magnitude $H = 22$ would have an apparent magnitude at the detection threshold $V_{\text{limit}} = 22$. The dashed lines show visibility constraints imposed by the minimum solar elongation angle $\gamma \geq 45°$.

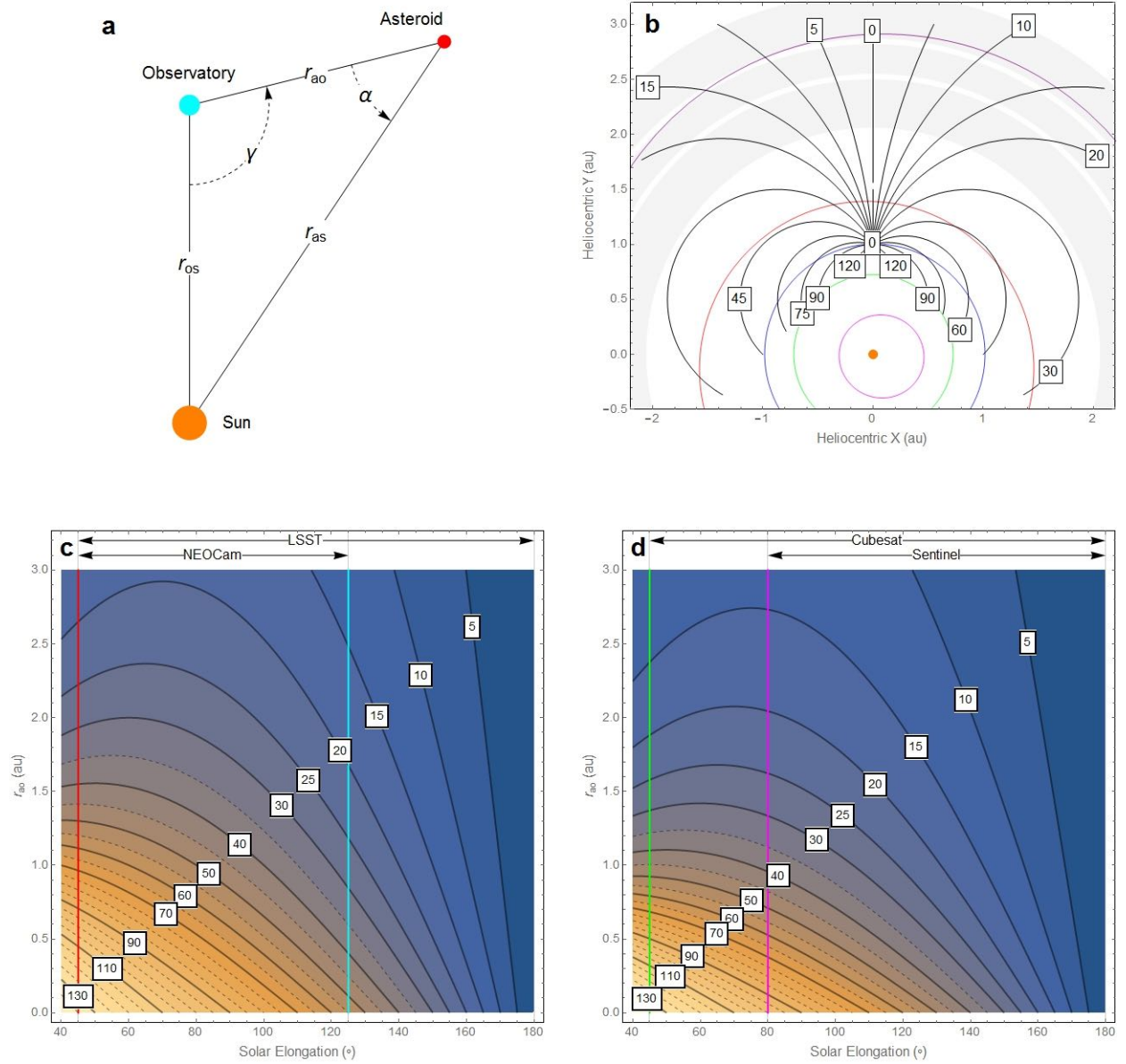

**Fig. 4.** Phase angle $\alpha$ and solar elongation $\gamma$ are shown in panel **a**, which illustrates Eq. (7). Panel **b** shows contours of constant phase angle for asteroids at various locations, assuming the telescope is at $(x, y) = (0, 1)$ in au. Colored rings represent the orbits of the planets. Panel **c** shows the contours of constant phase angle as a function of solar elongation $\gamma$ and the asteroid to observatory distance $r_{ao}$ for LSST and NEOCam with $(x, y) = (0, 1)$, and **d** shows phase angle for Sentinel and the Cubesat constellation, in their Venus-like orbits (*i.e.*, $(x, y) \approx (0, 0.71)$ ). Large phase angles such as $\alpha \geq 90°$ occur for only a small range of solar elongation, and have $r_{ao} < 1$.

## 2.3. Asteroids in Thermal IR

Among the many asteroid thermal models described in the literature, two seem best suited for this analysis. The FRM or Fast Rotating Model, also known as ILM (isothermal latitude model)(Harris and Lagerros, 2002; Harris, 2005; Kim et al., 2003; Lebofsky and Spencer, 1989; Mueller, 2012; Veeder et al., 1989) is the simplest of the three. I also employed NEATM—the Near Earth Asteroid Thermal Model (Delbó and Harris, 2002; Harris and Lagerros, 2002). And I derived a Generalized FRM (GFRM), as well as Tumble, which is a subset of GFRM for tumbling asteroids.

Each of these models can be put into a common framework, which assumes that incident radiation from the Sun is in equilibrium with emitted thermal radiation (Fig. 5). The asteroid is assumed to be spherical; $\theta$ and $\varphi$ are in conventional spherical coordinates ($\theta = 0°$ at the pole, $\theta = 90°$ at the equator). The absorption of radiation at a point is assumed to be Lambertian and is given by

$$\text{Absorbed}(\theta, \varphi) = \frac{S}{\sigma (5778\text{K})^4 \, r_{as}^2} \cos \Phi \int_0^\infty (1 - r_{dh}(\lambda)) B(5778\text{K}, \lambda) \, d\lambda \tag{8}$$

Where the effective black-body temperature of the Sun is 5778K, and $S = 1360.8 \, W/m^2$ is the solar constant, and $\Phi(\theta, \varphi)$ is the angle between the point $(\theta, \varphi)$ and the subsolar point (discussed more below).

$$B(T, \lambda) = \frac{2hc^2}{\left(e^{\frac{hc}{k_B T \lambda}} - 1\right) \lambda^5}, \tag{9}$$

is the Planck distribution. The quantity $r_{dh}(\lambda)$ is the directional-hemispherical reflectance in the notation of Hapke, 2012, which in general is a function of wavelength $\lambda$. It is called the point-albedo or normal-albedo in the notation of Lester et al., 1979.

Define the weighted average of an arbitrary function $g(\lambda)$ over the Planck distribution as

$$\langle g(\lambda)|T\rangle_B = \frac{1}{\sigma T^4} \int_0^\infty g(\lambda) B(T, \lambda) \, d\lambda, \tag{10}$$

The absorption equation (8) can then be rewritten as

$$\text{Absorbed}(\theta, \varphi) = \frac{S}{r_{as}^2} \left(1 - \langle r_{dh}(\lambda)|5778\text{K}\rangle_B\right) \cos \Phi. \tag{11}$$

The thermal radiation emitted at a point is then

$$\text{Emitted}(\theta, \varphi) = \eta \int_0^\infty \epsilon(\lambda) B(T(\theta, \varphi), \lambda) \, d\lambda, \tag{12}$$

where $\epsilon(\lambda)$ is the spectral emissivity (Hapke, 2012), and $\eta$ is a model-dependent parameter called the "beaming parameter" which will be discussed below.

Note that the NEATM radiative equilibrium occurs at a point. As a result, the relevant albedo (reflectivity) is $r_{dh}$, and *not* the geometric albedo $p$. The relevant emissivity is the spectral emissivity at a point, and *not* the bolometric emissivity (Hapke, 2012). This is relevant to the application of Kirchhoff's law, which must hold because the radiative equilibrium occurs at a point on the surface, so,

$$\epsilon(\lambda) = 1 - r_{dh}(\lambda), \tag{13}$$

which leads to

$$\text{Emitted}(\theta, \varphi) = \eta \, (\sigma \, T^4(\theta, \varphi) - \langle r_{dh}(\lambda) | T(\theta, \varphi) \rangle_B). \tag{14}$$

In the case of the NEATM, the radiative equilibrium is assumed to be instantaneous at each illuminated point on the asteroid, adjusted by

$$\text{Absorbed}(\theta, \varphi) = \text{Emitted}(\theta, \varphi), \tag{15}$$

which becomes

$$\frac{S}{r_{as}^2} (1 - \langle r_{dh}(\lambda) | 5778\text{K} \rangle_B) \cos \Phi = \eta \, (\sigma \, T^4(\theta, \varphi) - \langle r_{dh}(\lambda) | T(\theta, \varphi) \rangle_B). \tag{16}$$

In the case of FRM or GFRM, the assumption is that fast rotation and high thermal inertia force the balance to occur along infinitesimal rings of constant angle from the axis, as illustrated in Fig. 5. For each infinitesimal ring at angle $\theta$, the radiative equilibrium equation is

$$\int_{-\pi}^{\pi} \text{Absorbed}(\theta, \varphi) \sin \theta \, d\varphi = \text{Emitted}(\theta) = \int_0^\infty \int_{-\pi}^\pi \epsilon(\lambda) B(T(\theta), \lambda) \sin \theta \, d\varphi \, d\lambda, \tag{17}$$

which reduces to

$$\frac{S}{r_{as}^2} (1 - \langle r_{dh}(\lambda) | 5778\text{K} \rangle_B) \sin \theta = \pi \, (\sigma \, T^4(\theta) - \langle r_{dh}(\lambda) | T(\theta) \rangle_B), \tag{18}$$

in the FRM case where $\beta = 0$.

The treatment above differs from the usual derivation of asteroid thermal models because it is careful to use the proper albedo $r_{dh}$ and emissivity $\epsilon$. In other treatments the Bond albedo $A = pq$ appears instead of $\langle r_{dh}(\lambda) | 5778\text{K} \rangle_B$.

This substitution occurs through a separate approximation, over and above the point wise radiative equilibrium assumption, and is done to relate the models to observational evidence. Rather than observing the albedo (reflectance) $r_{dh}$ at a point on the surface, we instead observe the geometric albedo $p$ which is the albedo for a sphere (Hapke, 2012; Lester et al., 1979). In addition, we know that asteroids are not perfect Lambertian spheres, and have complex phase laws (3-5).

We can relate the model quantities to observed quantities by setting their Bond albedos to be equal to one another, $A_{\text{lambert}} = A_{\text{observed}}$. Since $A_{\text{lambert}} = r_{dh}$, we have

$$r_{dh}(\lambda) = p(\lambda) \, q \tag{19}$$

A further approximation that is frequently made is to assume that the geometric albedo $p(\lambda)$ is approximately equal to the observed visible light band geometric albedo $p_v$, i.e. $p(\lambda) \approx p_v$.

This is a rather poor approximation, as can be seen by calculating the fraction of the total incident solar flux within a wavelength band. A simple definition of the visual band is to use the response function

$$s_v(\lambda) = \begin{cases} 0 & \lambda < 0.4\ \mu m \\ 1 & 0.4\ \mu m \leq \lambda \leq 0.7\ \mu m. \\ 0 & \lambda > 0.7\ \mu m \end{cases} \qquad (20)$$

An alternative would be to define $s_v(\lambda)$ as the filter passband in a photometric filter system, including the sensor response.

The radiated energy flux within the visible band is $w_v = \langle s_v(\lambda)|5778K\rangle_B = 0.37$, so only 37% of the radiated flux is within the visual band. While the visible band has obvious relevance to observational data, it is not, by itself, a good estimate of the total energy.

There is more power in the near-IR. If we define near-IR by $0.7\ \mu m \leq \lambda \leq 2.4\ \mu m$, then $w_{\text{nir}} = 0.47$. The mid-IR (defined here as $2.4\ \mu m \leq \lambda \leq 10\ \mu m$) contribution is much smaller, with $w_{\text{mir}} = 0.037$.

A further complication is that observational measurements of absolute magnitude $H$, and by extension, geometric albedo $p_v$, are based on brightness within a band, which is determined by photon count flux, not energy flux.

The photon count equivalent of (10) is the weighted average

$$\langle g(\lambda)|T\rangle_N = \frac{h^3 c^2}{4\ k_B\ \zeta(3)T^3} \int_0^\infty g(\lambda) N(T,\lambda)\, d\lambda, \qquad (21)$$

Where $\zeta(s)$ is the Riemann zeta function, and the Planck distribution for photon count is

$$N(T,\lambda) = \frac{2c^2}{\left(e^{\frac{hc}{k_B T \lambda}} - 1\right)\lambda^4}. \qquad (22)$$

The distributions $B$ and $N$ are in general quite distinct with peaks at different wavelengths. In this terminology, the visible geometric albedo $p_v$ is given by

$$p_v = \langle p(\lambda)|5778K\rangle_N = \frac{1}{q}\langle r_{dh}(\lambda)|5778K\rangle_N \qquad (23)$$

The use of $p_v$ in asteroid thermal models amounts to an assumption that

$$\langle r_{dh}(\lambda)|5778\text{K}\rangle_B \approx \langle r_{dh}(\lambda)|5778\text{K}\rangle_N. \tag{24}$$

In general, the two sides of (24) are equal only if $r_{dh}(\lambda) = p_v/q$ is a constant independent of wavelength $\lambda$, which amounts to assuming that asteroids are gray bodies with no color. This is certainly not the case in practice, much of our understanding of asteroids comes from classifying them by color (i.e. spectral response)(DeMeo et al., 2009). Exploring the potential errors in the approximation is a topic of ongoing research.

The emissivity terms $\langle r_{dh}(\lambda)|T(\theta,\varphi)\rangle_B$ and $\langle r_{dh}(\lambda)|T(\theta)\rangle_B$ in (16) and (18) are also weighted averages over a Planck distribution, but over a much different range of temperatures than (24). For an individual asteroid the temperature ranges across $0 \leq T(\theta,\varphi) \leq T_{ss}$, where $T_{ss}$ is the maximum temperature that occurs at the sub-solar point. For most asteroids that have been observed, $175\text{K} \leq T_{ss} \leq 350\text{K}$, with the lower end of the range for asteroids far from the Sun such as the Trojans, and the higher temperature typically for near Earth asteroids. In all cases weighted average samples $r_{dh}(\lambda)$ over a very different range of wavelengths than the solar radiation which is at 5778K.

Common practice in the literature is to assume that $\epsilon = 0.9$, which is equivalent to assuming that

$$0.1 \approx \langle r_{dh}(\lambda)|T(\theta,\varphi)\rangle_B \approx \langle r_{dh}(\lambda)|T(\theta)\rangle_B. \tag{25}$$

The quality of this approximation is unclear, particularly across asteroids that have $T_{ss}$ vary by a factor of two, yielding weighted averages over very different distributions. The simplest interpretation of (25) is that $r_{dh}(\lambda) \approx 0.1$ in the wavelength range $2.4\ \mu m \leq \lambda \leq 50\ \mu m$.

Thus the implicit assumption in most treatments of asteroid thermal models is that the albedo is approximately given by

$$r_{dh}(\lambda) \approx \begin{cases} p_v q & \lambda \leq 2.4\ \mu m \\ 0.1 & 2.4\ \mu m \leq \lambda \leq 50\ \mu m' \end{cases} \tag{26}$$

Note that the wavelength ranges in (26) are approximate. The rationale behind them is that solar radiation is still significant out to $2.4\ \mu m$, however one could use a different limit. Thermal emission at typical asteroid temperatures covers the range $2.4\ \mu m \leq \lambda \leq 50\ \mu m$. Asteroid thermal modeling papers typically do not contain an explicit assumption for albedo like (26) and instead implicitly just assume that $p_v q$ is an approximation to the Bond albedo at all wavelengths, and that $\epsilon = 0.9$. It is far from clear how valid the approximation (26) for observed asteroids, and it is a topic of ongoing research.

The poorly characterized approximations in (24-26) are moot in the NEATM model because the free parameter $\eta$, is used to fit the model to observational data. The values of $p_v q$ and $\epsilon$ are mathematically irrelevant because $\eta$ compensates for them. In the FRM model this is not the case; there is no extra free parameter, so the quality of the approximations is of greater consequence.

Previous NEO studies have taken two approaches, broadly speaking, in their treatment of IR albedo. One group of papers ("Group 1"), which includes the original first expositions of FRM and NEATM as well as more recent studies (Delbó and Harris, 2002; Harris and Lagerros, 2002; Harris, 2005; Kim et al., 2003; Lebofsky and Spencer, 1989; Mueller, 2012; Veeder et al., 1989), focuses exclusively on emitted thermal radiation and ignores reflected solar IR.

A second set of papers ("Group 2"), which includes (Grav et al., 2012, 2011a, 2011b; Mainzer et al., 2011b, 2011c, 2011d, 2011e, 2015, 2014a, 2014b, 2012a, 2012b, 2012c, 2011a; Masiero et al., 2014, 2012a, 2012b, 2011)explicitly introduce an IR geometric albedo $p_{\mathrm{ir}}$to model reflected solar IR. In the notation used here, they assume that

$$r_{dh}(\lambda) \approx \begin{cases} p_{\mathrm{v}}q & \lambda \leq 2.4\ \mu m \\ p_{\mathrm{ir}}q & 2.4\ \mu m \leq \lambda \leq 10\ \mu m \end{cases}, \tag{27}$$

Again, the cut off between the two albedos is not specified explicitly. Group 2 papers concerned with the WISE IR space telescope assume that $p_{\mathrm{ir}}q$ is valid in the range $2.9\ \mu m \leq \lambda \leq 5.2\ \mu m$ which covers the WISE W1 and W2 bands, but do not say whether that is a hard limit on the applicability, or whether it could hold over a larger range of wavelengths. Mainzer et al., 2015 appears to assume that the $p_{\mathrm{ir}}q$ assumption is valid across $4\ \mu m \leq \lambda \leq 10\ \mu m$, which spans the two NEOCam bands.

Unfortunately Group 2 papers treat the Bond albedo $A$ and $\epsilon$ in the same way as those in Group 1. They assume that $A = p_{\mathrm{v}}q$, and $\epsilon = 0.9$. This is inconsistent with (27), because Kirchhoff's law holds for the NEATM model radiative equilibrium, as discussed above. It is inconsistent to assume that $r_{dh}(\lambda) \approx p_{\mathrm{ir}}q$ and $r_{dh}(\lambda) \approx 0.1$ for $2.4\ \mu m \leq \lambda \leq 10\ \mu m$. This is particularly true given that they obtain a very wide range of IR albedo $0.05 \leq p_{\mathrm{ir}} \leq 0.736$. Note that in this situation the free parameter $\eta$ cannot play the same compensatory role as in normal NEATM modeling because $\eta$ does not appear in the reflected light term.

At the very least this shows that comparison of IR telescopes requires careful matching of assumptions for the asteroid thermal models. Even studies that say they use the NEATM model may be using it differently because of assumptions about $p_{\mathrm{ir}}$.

In this study, I assume that Kirchhoff's law holds, and that phase integral $q$ is given by $q_{\mathrm{HG}}(G)$ in equation (4) for both the IR and visible bands, which leads to.

$$\begin{aligned} \epsilon &= 1 - p_{\mathrm{ir}}q \\ A &= p_{\mathrm{v}}q(1 - w_{\mathrm{mir}}) + w_{\mathrm{mir}}p_{\mathrm{ir}}q \end{aligned} \tag{28}$$

The result of the simple asteroid thermal models is to assign a temperature $T(\theta, \varphi, r_{\mathrm{as}}, r_{dh}, \epsilon)$, at each point on the model sphere. If we assume that $r_{dh}$ and $\epsilon$ are independent of wavelength $\lambda$, and make the assumptions (23) through (25) as is conventionally done, then the sub-solar point has the peak temperature

$$T_{\mathrm{ss}}(r_{\mathrm{as}}, A, \eta, \epsilon) = \left(\frac{S(1-A)}{r_{\mathrm{as}}^2 \epsilon \eta \sigma}\right)^{\frac{1}{4}} = 393.6\left(\frac{1-A}{r_{\mathrm{as}}^2 \epsilon \eta}\right)^{\frac{1}{4}}, \tag{29}$$

where $\epsilon$ is the emissivity, and $\sigma$ is the Stefan-Boltzmann constant. The parameter $\eta$ is model dependent; in FRM and GFRM, $\eta = \pi$. NEATM treats $\eta$ as an asteroid-dependent fitting parameter, typically $0.66 \leq \eta \leq 2.24$. In Tumble, $\eta = 4$.

For NEATM, (16) reduces to

$$T_{\text{NEATM}}(\theta, \varphi, r_{\text{as}}, A, \epsilon) = T_{\text{ss}}(r_{\text{as}}, A, \eta, \epsilon) \, (\sin \theta \cos \varphi)^{\frac{1}{4}}. \tag{30}$$

By definition, FRM requires that the axis of rotation is perpendicular to the plane defined by the telescope, asteroid, and the Sun—*i.e.*, the angle $\beta$ in Fig. 5a must be 90°. In that case,

$$T_{\text{FRM}}(\theta, \varphi, r_{\text{as}}, A, \epsilon) = T_{\text{ss}}(r_{\text{as}}, A, \pi, \epsilon) \, (\sin \theta)^{\frac{1}{4}} \tag{31}$$

$T_{\text{FRM}}$ ranges from 165 K to 360 K for typical values of $A$ and $0.7 \leq r_{\text{as}} \leq 3$ au.

It is believed that NEO spin axes are either isotropically distributed, or at least widely distributed (Bowell et al., 2014; Delbó and Viikinkoski, 2015; Magnusson, 1986), but definitive axis measurements for NEO are scarce. Generalized FRM drops the requirement that $\beta = 90°$. In GFRM,

$$T_{\text{GFRM}}(\theta, \varphi, r_{\text{as}}, A, \beta, \varphi_{\text{ss}}, \epsilon) = T_{\text{ss}}(r_{\text{as}}, A, \pi, \epsilon) \left( \int_{-\pi}^{\pi} \max(0, \cos \Phi(\theta, \varphi, \beta, \varphi_{\text{ss}})) \, d\varphi \right)^{\frac{1}{4}} \tag{32}$$

$$\Phi(\theta, \varphi, \beta, \varphi_{\text{ss}}) = \text{hav}^{-1}\big(\text{hav}(\theta - \beta) + \sin \theta \sin \beta \, \text{hav}(\varphi - \varphi_{\text{ss}})\big), \tag{33}$$

where the Haversine function $\text{hav}\,\theta = \sin^2\left(\frac{\theta}{2}\right)$, and $\theta_{\text{ss}} = \beta$ and $\varphi_{\text{ss}}$ are the angular coordinates of the sub-solar point.

It has been recognized that some asteroids, known as tumblers, may not rotate about a single axis, but instead may have multiple axes of rotation (Ďurech et al., 2005; Harris, 1994; Henych and Pravec, 2013; Warner et al., 2009). It is not known how many asteroids have this property, but some have been clearly identified by radar and optical observations (Harris, 1994; Henych and Pravec, 2013; Pravec et al., 2002).

Using the same fast rotation speed and high thermal inertia limit as with FRM, Tumble models a tumbling asteroid as having an average temperature over the entire surface with no angular dependence.

$$T_{\text{Tumble}}(\theta, \varphi, r_{\text{as}}, A, \epsilon) = T_{\text{ss}}(r_{\text{as}}, A, 4, \epsilon) \tag{34}$$

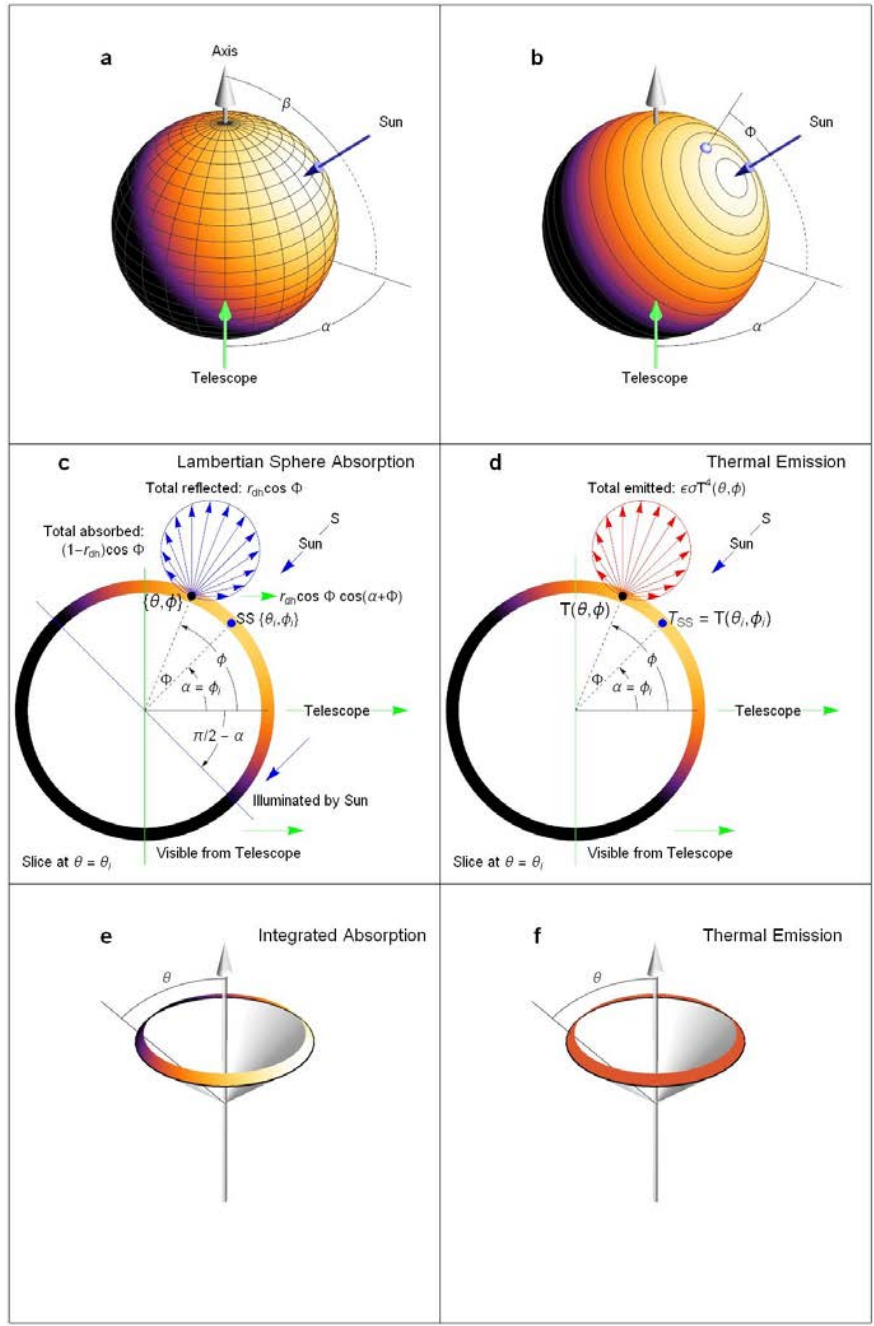

**Fig. 5.** Asteroid thermal models assume equilibrium between absorbed sunlight and emitted thermal radiation. Panel **a** shows the asteroid with incident sunlight at an angle $\beta$ from its axis of rotation. The intensity of the incident solar radiation is proportional to $\cos \Phi$, where $\Phi$ is the angle from the sub-solar point (panel **b**); causing $(1 - r_{dh})\cos \Phi$ to be absorbed (panel **c**), where $r_{dh}$ is the directional-hemispherical reflectivity. The temperature $T(\theta, \varphi)$ at a point leads to emission of thermal radiation (panel **d**). In the case of NEATM, the point-wise absorption and emission are assumed to be equal. FRM and GFRM models assume that fast rotation and high thermal inertia create an energy balance such that the integrated absorbed radiation on a ring of constant latitude (panel **e**) is balanced by emission a constant temperature $T(\theta)$ across the ring (panel **f**). The FRM model requires $\beta = 90°$; GFRM allows arbitrary $\beta$. The intensity distribution is visualized here with an arbitrary mapping of color to temperature.

is the lower bound on $T_{ss}$ for very fast tumbling about multiple axes. Real tumbling cases might range from a slow precession of the rotational axis, which would look much like GFRM cases, to the Tumble limit of fast tumbling. Over this range of tumblers, $\pi \leq \eta \leq 4$.

NEATM assigns zero temperature to the face of the asteroid away from the Sun. The parameter $\eta$ is called the "beaming" parameter, because it was originally was introduced to model excess radiation beamed into space by surface roughness and other optical effects.

Each of these thermal models is tailored to a specific limiting case of physical properties, and as such is likely to match some fraction of the NEO population. But no single model captures the diversity of asteroids that NEO searches aim to detect. NEO searches have the goal of cataloging a high fraction of the total, so they must cope with the diversity of asteroids, rather than seek a single average or typical approach.

Fig. 6 illustrates example cases of the temperature distribution for FRM, GFRM, NEATM, and Tumble thermal models. FRM and Tumble show no dependence on the phase angle $\alpha$, whereas GFRM and NEATM do.

The flux of thermal photons at the telescope aperture by photon count is given by

$$M_{\text{emit}}(d, G, p_v, p_{\text{ir}}, r_{\text{ao}}, r_{\text{as}}, \alpha) =$$

$$\frac{m^4 a^2 d^2 (1 - p_{\text{ir}})}{4 \, r_{\text{ao}}^2} \int_0^\infty \int_{\alpha - \pi/2}^{\alpha + \pi/2} \int_0^\pi s(\lambda) N(T(\theta, \varphi, r_{\text{as}}, A(G, p_v, p_{\text{ir}}), 1 - p_{\text{ir}}), \lambda) \sin^2 \theta \cos(\alpha - \varphi) d\theta \, d\varphi \, d\lambda,$$

(35)

Where $m = 0.97$ is the reflectivity of each of the four mirrors in both Sentinel and NEOCam, the aperture $a = 0.5$ m, $s(\lambda)$ is the sensor response function, and $N(T, \lambda)$ is the Planck photon count flux (22).

The flux by power rather than photon count, $F_{\text{emit}}(r_{\text{ao}}, r_{\text{as}}, d, G, p_v, p_{\text{ir}}, \alpha)$, is obtained from Eq. (35) by replacing $N(T, \lambda)$ with $B(T, \lambda)$ from Eq. (18).

Because NEATM model is illuminated only on the Sun-ward facing side (Fig. 6) and the phase angle $\alpha$ dependence is effectively the same as in the optical case, Eq. (35) effectively includes within it the phase function $\psi(\alpha)$ for a Lambertian sphere, which is obtained by integrating the term $\sin^2 \theta \cos(\alpha - \varphi)$ over $0 \leq \theta \leq \pi$, $\alpha - \pi/2 \leq \varphi \leq \alpha + \pi/2$.

Indeed, a potentially serious limitation of all the thermal models discussed here is that they assume that the asteroid is a perfect Lambertian sphere—an assumption that spacecraft images have shown to be an oversimplification. The differences between real asteroid shapes and Lambertian spheres has implications for modeling the performance of an NEO survey telescope, particularly in the IR. For observations made at phase angle $\alpha$, complex functions such as those in Eqs. (3) and (5) are used to model the visible band phase function $\psi(\alpha)$; in contrast, the corresponding functions to model IR emission take the same asteroid to behave like a smooth sphere.

This is problematic because the energy that powers IR emission depends on solar illumination. When NEATM assumes a point-by-point energy balance, those points are the same ones reflecting in complex ways modeled by complex forms of $\psi(\alpha)$. FRM also assumes a Lambertian sphere, although it introduces a smearing by rotation that makes this approximation somewhat more plausible.

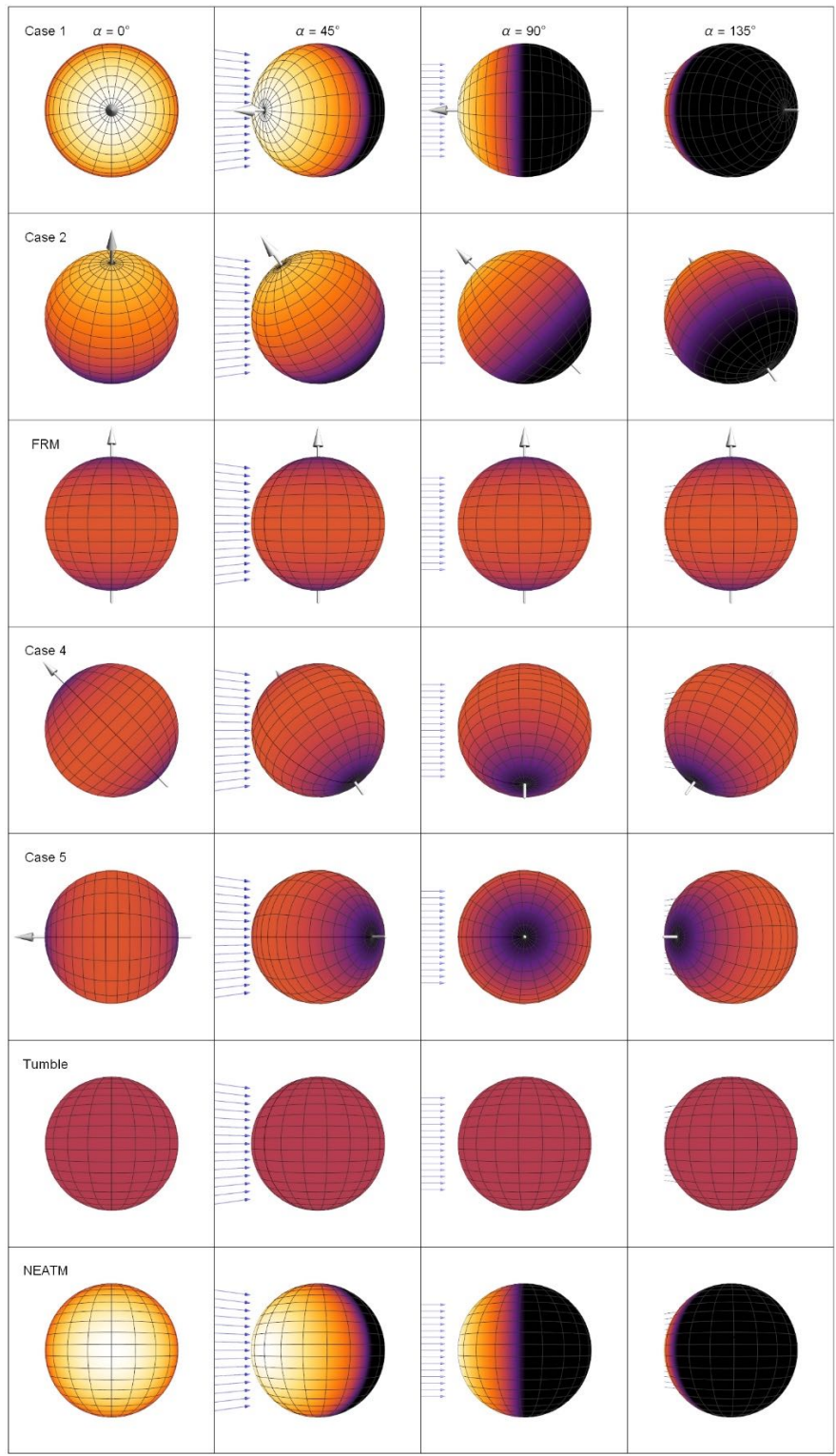

**Fig. 6.** Examples cases for FRM, GFRM, Tumble and NEATM. Cases are in rows; phase angle $\alpha$ variations for each case are in columns. In the $\alpha = 0$ column light from the Sun goes directly into the page. In other cases the Sun direction is shown with blue arrows; solar angle $\beta = 0$ in all cases. No spin axis is shown for Tumble and NEATM because these models are not dependent upon it.

More accurate IR models would, like $\psi(\alpha)$, account for the fact that real asteroids have irregular shapes as well as craters, boulders, and other terrain roughness. Thermal IR emissions from asteroids are also affected by coherent scattering in the regolith due to multiple scattering and surface roughness, shadow hiding, and many other effects (Jian-Yang Li, 2005; La Forgia F., 2014; Muinonen et al., 2012, 2010a, 2010b; Penttilä, 2011; Shkuratov et al., 2013; Videen and Muinonen, 2015). The effects will not be the same as for reflected light because thermal IR is emitted from the surface in all directions, whereas visible light is incident from a single direction (*i.e.,* the Sun). Recent work has focused on making thermal models for asteroids that include surface roughness, multiple scattering, and other effects (Davidsson and Rickman, 2014; Björn J. R. Davidsson et al., 2015; Maclennan and Emery, 2014; Rozitis and Green, 2014, 2011). While these models are a step in the right direction, the current state of the art does not reconcile visible light curves with those in the thermal IR.

Thermophysical modeling of heat transfer from actual asteroid shapes may offer the best approach for studying individual asteroids (Delbó and Tanga, 2009; Hanuš et al., 2015; Lagerros, 1996; Mueller et al., 2005, 2004; Müller, 2002). To compare telescopes, however, we need to construct statistical distributions of hypothetical asteroids. Whatever parameters are required to characterize an asteroid must be well enough known for a set of asteroids that that we can generalize to a hypothetical population. Observations of the detailed irregular shapes of a handful of asteroids are not sufficient for such generalization.

Although the HG phase function (3) is not ideal, it does encapsulate the observed departures from a Lambertian sphere into a single parameter $G$, which has been measured on enough asteroids that we understand something of its statistical distribution. In the context of NEO search comparison, any improvement on $\psi(\alpha)$ and its connection to IR models should also summarize properties in a small number of parameters that are well enough known to construct statistical distributions.

In the meantime, Lambertian models are widely used (Harris and Lagerros, 2002; Harris, 2005; Kim et al., 2003; Lagerros, 1998, 1997, 1996; Rozitis and Green, 2011; Wolters and Green, 2009)

## 2.4. Asteroids in Reflected IR

I model the reflected solar flux falling within the thermal IR band as

$$M_{\text{ref}}(d, G, p_{\text{ir}}, r_{\text{ao}}, r_{\text{as}}, \alpha) = \frac{m^4 a^2 d^2 p_{\text{ir}} \psi(G,\alpha)}{r_{\text{ao}}^2} \int_0^\infty s(\lambda) N(5778 \text{ K}, \lambda) \, d\lambda. \quad (36)$$

The total IR light observed by the telescope is

$$M_{\text{total}}(d, G, p_{\text{v}}, p_{\text{ir}}, r_{\text{ao}}, r_{\text{as}}, \alpha) = M_{\text{ref}}(d, G, p_{\text{ir}}, r_{\text{ao}}, r_{\text{as}}, \alpha) + M_{\text{emit}}(d, G, p_{\text{v}}, p_{\text{ir}}, r_{\text{ao}}, r_{\text{as}}, \alpha). \quad (37)$$

## 2.5. Telescope Capture Ratios

A thermal IR telescope captures only a portion of the IR spectrum. We can define the capture ratio, for both power $C_W$ and photon flux $C_N$, using

$$C_W(r_{as}, A, d) = \frac{F_{telescope}(d, G, p_v, p_{ir}, r_{ao}, r_{as}, \alpha)}{F_{ideal}(d, G, p_v, p_{ir}, r_{ao}, r_{as}, \alpha)}$$

$$C_N(r_{as}, A, d) = \frac{M_{telescope}(d, G, p_v, p_{ir}, r_{ao}, r_{as}, \alpha)}{M_{ideal}(d, G, p_v, p_{ir}, r_{ao}, r_{as}, \alpha)},$$

(38)

where $F_{telescope}$, $M_{telescope}$ are obtained from Eq. (35) using the sensor response function $s(\lambda)$ for the telescope in question, and $F_{ideal}$, $M_{ideal}$ have $s(\lambda) = 1$ (*i.e.,* they are equally sensitive to photons of all wavelengths).

In the case of Sentinel, $s(\lambda)$ is a notch from 5–10 μm, with a linear taper to 12 μm. NEOCam has $s(\lambda)$ with two notch passbands, 4–5.2 μm, and 6–10 μm. The capture ratio is a function of asteroid distance from the Sun $r_{as}$ because as the temperature of the asteroid changes, different amounts of the energy will be captured by the fixed sensor response.

## 2.6. Asteroid Scenarios

IR observation of an asteroid in Eq. (38) depends on physical parameters of the asteroid $d, G, p_v, p_{ir}$ and its observational geometry $r_{ao}, r_{as}, \alpha$. Visible band observation (Eq. (6)) is similar, but does not depend on $p_{ir}$.

As a simplification, much of the analysis here focuses on the asteroid size range $20 \leq d \leq 1000$ m. Other asteroid physical properties are evaluated using three specific scenarios: Low ($p_v = 0.05$, $G = 0.13$), Med ($p_v = 0.14$, $G = 0.25$), and High ($p_v = 0.46$, $G = 0.4$). I will refer to the Low scenario as "low-brightness", and so forth for the other cases. Low-brightness should be understood to be in the visible light band for asteroids of its size (i.e. low $H$ for a given $d$ as per equation (1)). Note that this is more than just low albedo $p_v$ – it is a combination of both $p_v$ and the slope parameter $G$. These scenarios are compatible with prior work (Morbidelli et al., 2002) and are intended to be illustrative of important cases. Other values are considered in the sensitivity analysis.

The IR albedo is assumed to be $p_{ir} = 1.27\, p_v$. In the sensitivity analysis, $0.5 \leq p_{ir}/p_v \leq 2.0$.

## 2.7. Terrestrial Effects

LSST is located on Earth and consequently must observe through the atmosphere. At least four important factors affect its performance as a result. The first is atmospheric extinction caused by light being absorbed in the atmosphere. The mass of the column of air that the telescope looks through at zenith is lower than the mass for a line of sight near the horizon. A second factor is astronomical seeing, a measure of distortion in the image caused by atmospheric turbulence. Seeing depends both on air mass and on the observatory site-specific statistical distribution of turbulence. The third effect is sky brightness, which provides background noise that can confound operations. The night sky is in general brighter at the horizon near sunset and sunrise, an effect that is important for observing objects near the Sun (*i.e.,* having low solar elongation angle $\gamma$). The moon also contributes to sky brightness, and its position and phase change constantly. The fourth effect arises from the telescope itself rather than the atmosphere. As the telescope points at different angles relative to the zenith, its mechanical structure and optical elements experience different gravitational forces, which can cause mechanical flexing that distorts the image.

For LSST, each of these effects has been either calculated by the LSST team (Ivezić and Council, 2011; Ivezić et al., 2008) or measured at the Cerro Pachón site where LSST is being built (Delgado et al., 2014). The characterization work include atmospheric radiative transfer models such as MODTRAN (Acharya et al., 1999; Berk et al., 1999; Postylyakov, 2004) and empirical measurements of sky brightness by other astronomical observatories in Chile (Jones et al., 2014; Krisciunas et al., 2007; Patat, 2008; Patat et al., 2010).

Unfortunately, there is no simple relationship between heliocentric coordinates $(x, y)$ of an object and the solar elongation $\gamma$ and the elevation angle above the horizon at which the survey will observe the object. In general, objects having small values of $\gamma$ must by geometry be observed at low elevation angles. As an example, an object near the ecliptic at $\gamma = 45°$, will be approximately 27° above the horizon at the end of astronomical twilight (defined as the Sun being 18° below the horizon), depending on observatory latitude and the time of year. Such objects are high in the sky only during daylight hours.

As the value of $\gamma$ increases, so does the range of elevation angles at which it may be observed. The observing cadence is a complicated function of the strategic objectives of the survey, the quality objectives of observing at low air mass, and tactical issues, such as ensuring that the field of view remains a distance from the moon that depends on lunar phase.

I obtained the modeled limiting magnitude at $SNR = 5$ detection threshold from the LSST operational simulator (Delgado et al., 2014) for 2.5 million simulated observations, equivalent to each exposure taken by the telescope during its 10 year initial observing survey, with all of the relevant terrestrial observational effects taken into account. These were binned by solar elongation angle $\gamma$ to obtain a statistical distribution of limiting magnitudes as a function of $\gamma$, as summarized in Fig. 7. The median value is used in most calculations in this study, with other cases considered in the sensitivity analysis.

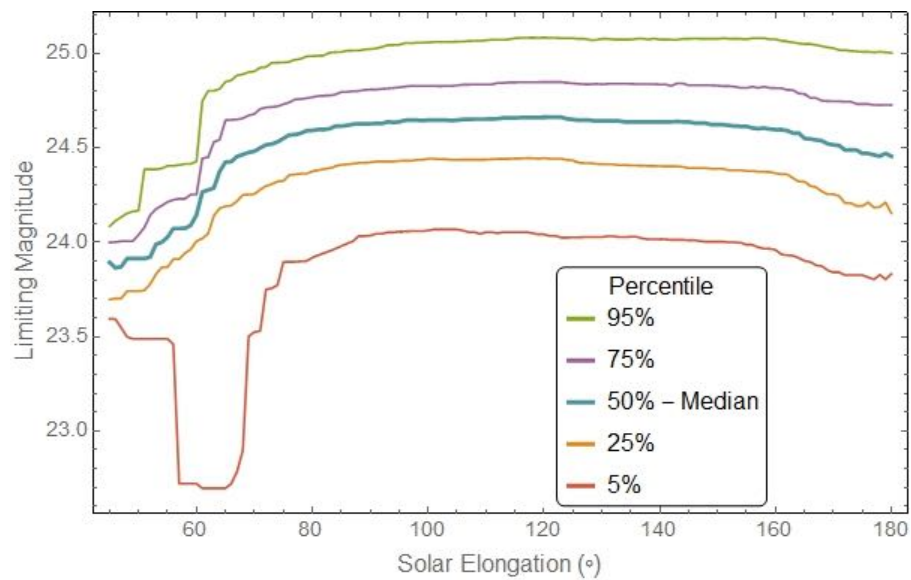

**Fig. 7.** *Summary of LSST limiting magnitude for 2.5 million simulated observations over 10 years.*

## 2.8. Search Volumes

For a given telescope and source object, one can define the region in space where the object is observable—this is the search volume. The search volume is constrained by the field of regard ($FOR$) of the telescope, which spans the portion of the sky that the instrument can observe. $FOR$ is limited by both Sun avoidance and aspects of the telescope design; it can be expressed in terms of the range of solar elongation angles $\gamma$. For LSST, the $FOR$ spans $\gamma \geq 45°$; for Sentinel, the $FOR$ is $\gamma \geq 80°$. Each of the five Cubesat satellites would have $\gamma \geq 45°$. NEOCam has an unusual $FOR$ of $45° \leq \gamma \leq 125°$.

The $FOR$ is an angular constraint of infinite extent. The search volume is the finite zone inside the $FOR$ within which an asteroid of particular properties (*i.e.*, $d, G, p_v, p_{ir}$) can be detected by the telescope to $SNR \geq 5$. Inclusion in the search volume means that the asteroid would be detected if the telescope is pointing in a field of view ($FOV$) around the line of sight (LOS) that contains it. However, each telescope has a field of view (FOV) that is much smaller than its $FOR$, *i.e.*, $FOV \ll FOR$.

A search strategy adopts an observing schedule and cadence that samples within the $FOR$. The cadence is determined by the exposure time spent observing at each point, by the slew rate moving between observing positions, and by other observational considerations, such as avoiding the moon or giving more coverage to areas of the sky that are more likely to contain asteroids (*e.g.*, the plane of the ecliptic). Because $FOV \ll FOR$, the search volume overstates the actual search performance of the telescope—due to finite exposure times and the reality of the cadence, the telescope is unlikely to detect every potentially visible asteroid. Nevertheless, the search volume imposes an important upper bound constraint on the telescope performance.

I compute the visibility from $H$ and the chosen phase angle law and parameters, as with Fig. 3. Eq. (6) gives the implicit equation for the search volume, where the limiting apparent magnitude $V_{\text{limit}}$ for LSST is given by the median curve in Fig. 7, and $V_{\text{limit}} = 20.58$ for the Cubesat-5 constellation.

In heliocentric coordinates for the asteroid $x_a, y_a$ and observatory $x_o, y_o$, the separation distances

$$r_{ao} = \sqrt{(x_a - x_o)^2 + (y_a - y_o)^2}$$
$$r_{as} = \sqrt{x_a^2 + y_a^2}$$
$$r_{os} = \sqrt{x_o^2 + y_o^2}.$$

(39)

Eq. (6), with substitutions from Eq. (39), can then be solved numerically to find the search volume for visible band telescopes.

In the IR, the criterion for visibility at a given $SNR$ is

$$t\, M_{\text{total}}(d, G, p_v, p_{ir}, r_{ao}, r_{as}, \alpha) \geq SNR \sqrt{t\, M_{\text{noise}}}\,, \tag{40}$$

where $M_{\text{total}}$ is given by Eq. (37), $M_{\text{noise}}$ is the noise floor, and $t$ is the exposure time.

Because Sentinel and NEOCam both use low-noise sensors of similar technology, and both operate at 35 K, the dominant noise source for these systems would be IR emission from zodiacal dust (Leinert et al., 1998; Mainzer et al., 2015). The density and temperature of the zodiacal dust varies both with ecliptic latitude and direction, complicating the modeling of its effects. I did not compare the zodiacal dust models used by each of the teams to check for consistency. A better approach for future detailed simulations would adopt a consistent model and include the variations.

I simplify the effects of noise by using the worst-case estimates within the plane of the ecliptic, which is $M_{\text{noise}} = 12{,}370$ for Sentinel and $M_{\text{noise}} = 10{,}000$ for NEOCam. In the best case, away from the ecliptic, the estimated noise levels are $M_{\text{noise}} = 5000$ for Sentinel (Harold Reitsma personal communication, 2015) and $M_{\text{noise}} = 2000$ for NEOCam (Amy Mainzer, personal communication, 2015). Sensitivity to variations in these thresholds are calculated.

For both telescopes, $SNR = 5$ and $t = 180$. As with the calculations described above for the visual-band telescopes, the implicit Eq. (6) was solved numerically in the plane of the ecliptic to obtain search volumes.

### 2.9. Visibility Ratios

Plotting the search volumes is helpful in building intuition, but it is hard to compare the irregularly shaped search volumes. To facilitate comparison, I use the visibility ratio, a metric that captures the breadth and the depth of the search volumes.

Consider a spherical shell centered on the Sun with radius $r$. As NEOs move on their orbits they will at some point cross the shell. A metric of the search breadth at radius $r$ is to take the ratio of the area of the shell that is inside the search volume to that which is outside. A two-dimensional version of this metric is plotted in Fig. 8, which shows the arc length in the plane of the ecliptic within the search volume.

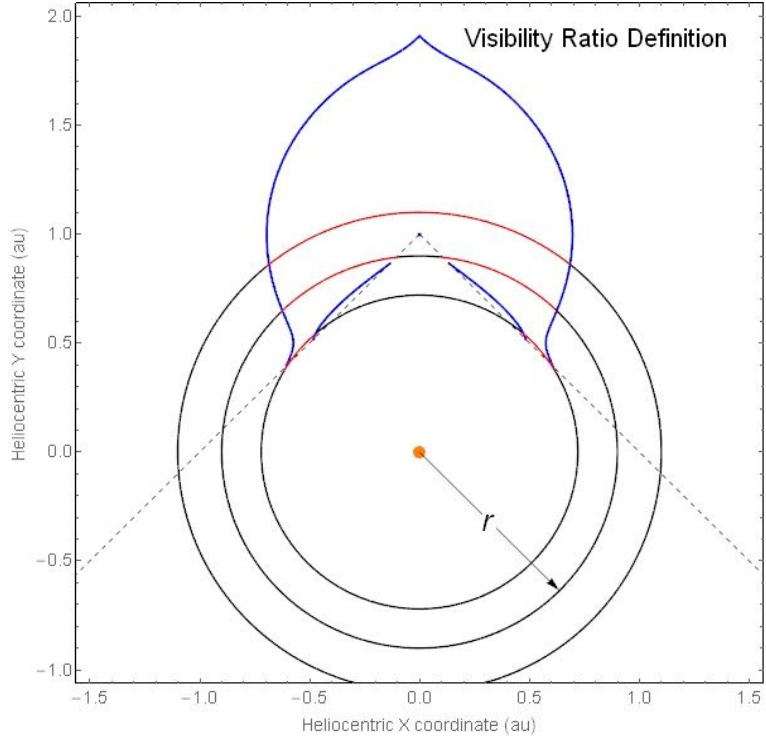

**Fig. 8.** Schematic representation of a visibility ratio. A sample search volume is shown in blue. For any given heliocentric radius r, we can define the portion of the arc that lies inside the search volume (shown in red), and the portion outside (shown in black). The ratio of the red to black arc length (in two dimensions) or surface area (in three dimensions) is the radial visibility ratio.

For visual-band telescopes, the visibility ratio is obtained by integrating Eq. (6) in the following manner, using Eqs. (39) and (42).

$$S_\text{v}(d, G, p_\text{v}, r) = \frac{1}{2\pi} \int_0^{2\pi} h\left(V(d, G, p_\text{v}, r_\text{as}, r_\text{ao}, \alpha) - V_\text{limit}\right) d\omega \qquad (41)$$

$$(x_\text{a}, y_\text{a}) = (r\cos\omega, r\sin\omega), \qquad (42)$$

where $h$ is the unit step function

$$h(x) = \begin{cases} 0 & x \leq 0 \\ 1 & x > 0. \end{cases}$$

In the IR band, the corresponding integral is

$$S_\text{ir}(d, G, p_\text{v}, p_\text{ir}, r) = \frac{1}{2\pi} \int_0^{2\pi} h\left(t\, M_\text{total}(d, G, p_\text{v}, p_\text{ir}, r_\text{ao}, r_\text{as}, \alpha) - SNR\sqrt{t\, M_\text{noise}}\right) d\omega, \qquad (43)$$

with the same reliance on Eqs. (39) and (42). Both the IR and visual band integrals can be evaluated numerically using standard techniques.

## 2.10. Fraction of NEOs in Search volume

Given the visibility ratio $S$ and the NEO population radial probability $P(r)$ we can find the fraction of the NEO population within the search volume at any moment in time, for both visible-band and infrared observations:

$$f_v(d, G, p_v) = \int_0^\infty P(r)\, S_v(d, G, p_v, r)\, dr$$
$$f_{ir}(d, G, p_v, p_{ir}) = \int_0^\infty P(r)\, S_{ir}(d, G, p_v, p_{ir}, r)\, dr. \qquad (44)$$

Under the assumption that $P(r)$ is stationary, we would expect that $f$ would be nearly constant in time as asteroids move in and out of the search volume.

## 2.11. Integrated NEO Fraction

The number of NEOs of a given absolute magnitude $H$ can be estimated (Morbidelli et al., 2002) (Bottke et al., 2002) as

$$Q(H) = 13.26\ 10^{0.35(H-13)}. \qquad (45)$$

This corresponds to a distribution with respect to size that is proportional to $d^{-1.75}$. We can use this relation to find the total fractions of asteroids from $140 \leq d \leq 1000$, overall and separately in visible and IR bands, for a given set of parameters $G, p_v, p_{ir}$:

$$\Gamma_o(p_v) = \int_{140}^{1\,000} Q(H(x, p_v))\, dx$$

$$\Gamma_v(G, p_v) = \frac{1}{\Gamma_o(p_v)} \int_{140}^{1\,000} f_v(x, G, p_v)\, Q(H(x, p_v))\, dx \qquad (46)$$

$$\Gamma_{ir}(G, p_v, p_{ir}) = \frac{1}{\Gamma_o(p_v)} \int_{140}^{1\,000} f_{ir}(x, G, p_v, p_{ir})\, Q(H(x, p_v))\, dx.$$

## 3. Results

### 3.1. IR Sensitivity & Thermal Models

The ratio of the total energy radiated as IR (both reflected and thermally emitted) to the total energy of reflected light in the visible band is

$$IRtoV = \frac{1 - q_{\text{HG}}(G)\left((1-w)\,p_v - w\,p_{\text{ir}}\right)}{q_{\text{HG}}(G)\,(1-w)\,p_v}, \qquad (47)$$

Where $w$ is the fraction of the solar energy in the IR band (i.e. $w = \langle s_{\text{ir}}(\lambda)|5778\text{K}\rangle_B$; $w = 0.280$ if $\lambda_{\text{low}} = 1\,\mu m$, $\lambda_{\text{high}} = \infty$.)

Using the typical range of values for $G$, $p_v$, and $p_{\text{ir}}$, $4.7 \leq IRtoV \leq 197$—asteroids emit much more energy in IR than in the visual band. This is a primary motivation for searching for NEOs with IR telescopes, and it helps to explain why it might be possible for a 0.5 meter aperture telescope like Sentinel or NEOCam to have good performance.

Unfortunately, $IRtoV$ is not the only important factor. Although asteroids typically radiate more energy in the IR, the passbands of Sentinel and NEOCam would allow them to capture only a fraction of the IR incident on their sensors. The expected capture ratios for these instruments are illustrated in Fig. 9.

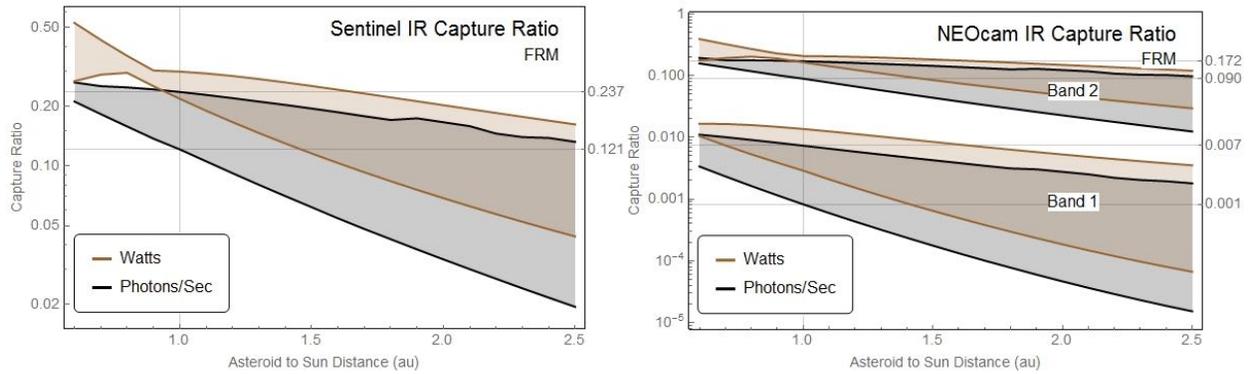

**Fig. 9.** Infrared capture ratios, $C_W$ in power (brown) and $C_N$ in photon flux (black), for Sentinel and NEOCam as a function of asteroid distance from the Sun. NEOCam has two IR sensor bands, which are plotted separately. Shaded regions indicate the results of varying the parameters through ranges: $0.13 \leq G \leq 0.4$, $0.02 \leq p_v \leq 0.46$, and $0.5 \leq p_{\text{ir}}/p_v \leq 2.0$.

For an asteroid located in Sentinel's prime working region near 1 au, the instrument will capture between 12% and 24% of the IR photon flux arriving from the asteroid, depending on its physical parameters. At a distance of 2.5 au, the capture ratio falls to between 2% and 12% of the incident IR photon flux. NEOCam does slightly less well, capturing between 9% and 17% of the IR photon flux at 1 au in band 2, dropping to 1% to 9% of the photon flux at 2.5 au. Band 1 captures very little of the photon flux: 0.01% to 0.07% at 1AU.

The passbands of both IR telescopes are narrow compared to the bandwidth of the source radiation for the objects in the region of interest. This decreases the effect of Eq. (47) and removes some of the IR advantage, with a strong effect on the $SNR$ and other properties, as shown below.

NEOCam has two sensor arrays that use distinct passbands: band 1 (4–5.2 μm) and band 2 (6–10 μm). Because band 1 is narrow and most of the thermal IR radiated by asteroids has $\lambda > 5.2$ μm, this sensor is not useful on its own for detecting asteroids. When an asteroid is detected by the band 2 sensor at $SNR = 5$, the band 1 sensor will typically have an $SNR \approx 1.5$, such that it would be more effective to add the two bands than to consider each separately. The band 1 sensor would be expected to achieve $SNR = 5$ only for objects in a search volume that is typically 1% to 5% the size of the search volume for the two bands combined. Within that far smaller region, both bands could be used simultaneously to constrain estimates of asteroid spectra or to perform science goals other than detection.

Prior analyses by the Sentinel and the NEOCam groups used both FRM and NEATM, but the two groups used different methods. The Sentinel group's simulations (Buie and Reitsema, 2015; Lu et al., 2013) followed the general approach discussed in section 2.3 as Group 1: reflected IR was assumed to be zero, $p_{\text{ir}}$ was not explicitly modeled, and emissivity was considered constant at $\epsilon = 0.9$. The NEOCam group used the Group 2 approach (Mainzer et al., 2015), which explicitly assumes that $p_{\text{ir}} \neq 0.1$, yet also uses $\epsilon = 0.9$.

The top two panels of Fig. 10 illustrate the effects of these differing assumption on $T_{FRM}$. The correct $T_{FRM}$, plotted as a function of $p_{\text{v}}$ and $p_{\text{ir}}$ with correct treatment of $A$ and $\epsilon$, is shown in Fig. 10a. I assume here that the phase parameter $G = g(p_{\text{v}})$, but the results are very similar when $G = 0.15$ or other values. Fig. 10b shows $T_{FRM}$ under the assumptions on $A$ and $\epsilon$ used by both Group 1 and Group 2 studies; the assumptions have a pronounced effect on temperature. The same arbitrary mapping of color to temperature is used for both Figs. 10a and 10b.

Fig. 10c illustrates the effect of the Group 2 assumptions on photon flux by plotting the ratio of the flux as calculated using those assumptions to the flux calculated using the correct assumptions, for sample asteroid parameters. The ratio is plotted over the search volume for NEOCam. The ratio is greatest at the point in the search volume that is closest to the Sun (cyan arrow); its minimum value occurs at the point farthest from the Sun (magenta arrow). The plot for Sentinel (not shown) is qualitatively similar, and also has a maximum and minimum at the closest and farthest points from the Sun, respectively, within its search volume.

Figs. 10d and 10e demonstrate how, for both Sentinel (Fig. 10d) and NEOCam (Fig. 10e), the ratios of photon flux under the different assumptions vary with $p_{\text{v}}$ and $p_{\text{ir}}$ at the farthest point to the Sun in the search volume. Where the ratio is one (bold contour), the Group 1 (for Sentinel) or Group 2 (for NEOCam) results are equal to the correct value. The photon flux as calculated by the Group 2 / $\epsilon = 0.9$ approach gives an inaccurate answer everywhere else. The range of ratios for NEOCam is from 1.07 to 0.59, a factor of about 1.83 across $p_{\text{v}}$ and $p_{\text{ir}}$. For Sentinel the range is from 1.07 to 0.44, a factor of 2.44.

Based on the results shown in Fig. 10, I find that the different assumptions used in previous work by the Sentinel and NEOCam teams tends to misestimate the expected performance of these instruments, overestimating their performance in some cases by up to ~7%, but underestimating it in other cases by up to ~229%.

When $A$ and $\epsilon$ are treated correctly with respect to $p_{\text{ir}}$, there is a tradeoff between $p_{\text{v}}$ and $p_{\text{ir}}$ in the impact on the photon flux, and with it the $SNR$. Fig. 11 illustrates this tradeoff for a 140 m asteroid positioned at example points $(x, y)$ heliocentric coordinates chosen near the edge of the search volume. Because it is at the edge, the asteroid can be detected at $SNR \geq 5$ for some combinations of $p_{\text{v}}$ and $p_{\text{ir}}$ but not for others.

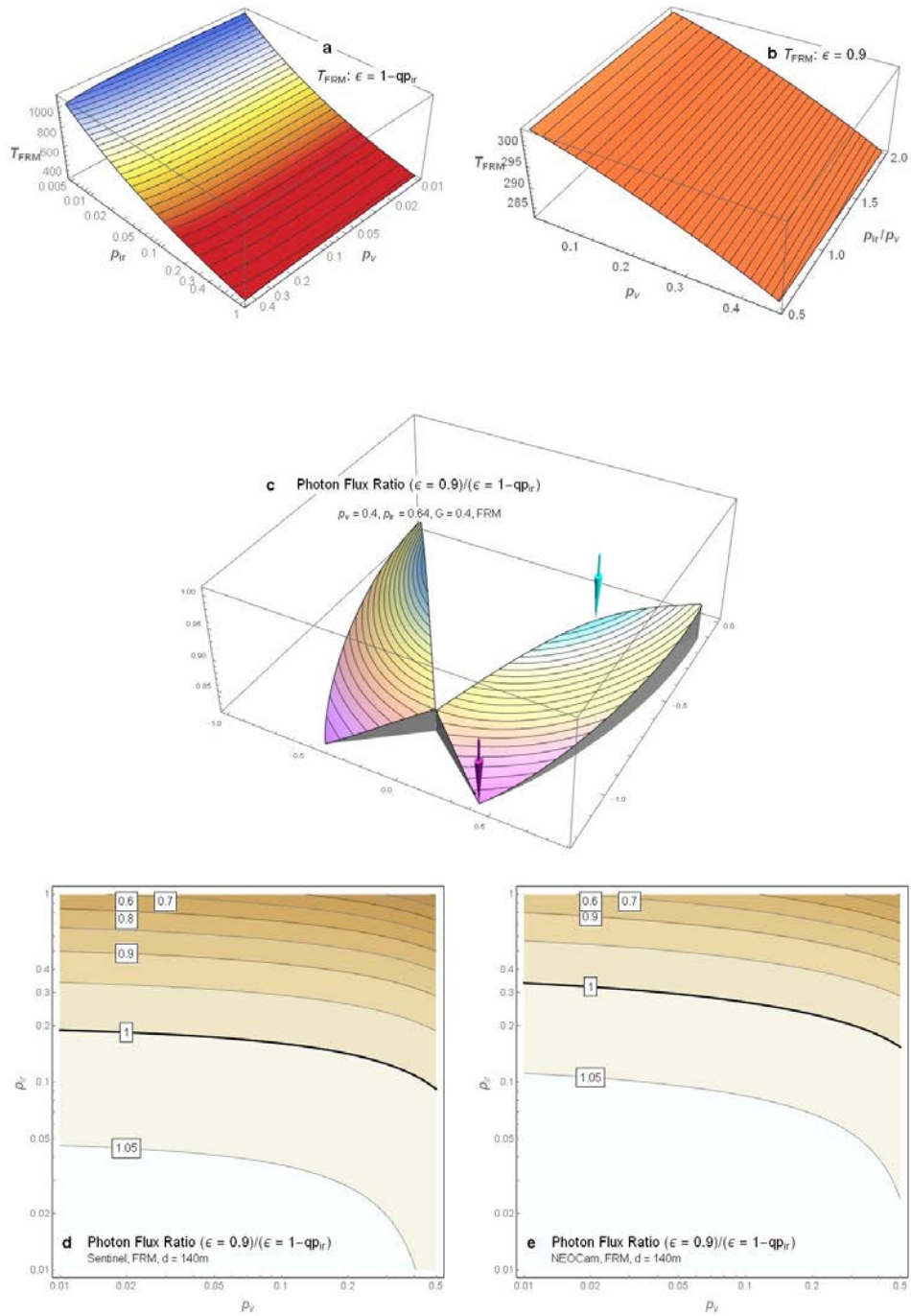

**Fig. 10.** Impact of assumptions on A and $\epsilon$ on temperature, photon flux, and SNR. Panel **a**: the effects of the visible-light geometric albedo $p_v$ and the IR albedo $p_{ir}$ on the FRM peak temperature $T_{FRM}$, evaluated at $r_{as} = 1$ au, using $G = g(p_v)$. Panel **b**: $T_{FRM}$ as calculated by Group 1 and Group 2 studies (see text) that fix $\epsilon = 0.9$ in violation of Kirchhoff's Law. Panel **c**: ratio, plotted over the search volume for NEOCam, of photon flux when $\epsilon = 0.9$ to photon flux when $\epsilon = 1 - qp_{ir}$; the ratio has a maximum at the point in the seach volume closest to the Sun (cyan arrow) and a minimum at the farthest point from the Sun (magenta arrow). Panels **d**, **e**: the photon flux ratios for Sentinel (**d**) and NEOCam (**e**) at the points in the search volume that are farthest from the Sun (magenta arrow in **c**).

## 3.2. Search volumes

Using Eqs. (6–8), (31), (39), and (40), I plotted the search volumes in both the visual and IR bands for the four telescopes. Fig. 12 and 13 plot the search volumes in two orthogonal planes for LSST and Sentinel. Figs. 14 shows the search volumes in the plane of the ecliptic are shown for LSST and NEOCam. Fig 15 does the same for the Cubesat-5 constellation.

For the visual-band telescopes LSST and Cubesat-5, the search volume is strongly dependent on the asteroid scenario. Those scenarios have less impact on the IR-band telescopes Sentinel and NEOCam. At $r_{as} = 1\ AU$, I find that $T_{FRM}(1, A_{High}) = 340.2$ K, $T_{FRM}(1, A_{Med}) = 305.1$ K, and $T_{FRM}(1, A_{Low}) = 299.0$ K (all with $p_{ir}/p_v = 1.27$). The temperature difference from Low to High is about 41 K.

Note that search volumes are three-dimensional; the plots here show the two-dimensional intersection between the region and the plane of the ecliptic (or, in the case of Fig. 13, a plane perpendicular to the ecliptic). For LSST and Cubesat-5, the three-dimensional search volume is bounded by a surface of revolution defined by rotating the region in the plane of the ecliptic around the axis from the telescope to the Sun. For the IR telescopes, the search volume is larger in directions away from the plane of the ecliptic because the noise floor $M_{noise}$ drops off rapidly with increasing ecliptic latitude, reaching a minimum value at 90° (Fig. 13). Although the search volume for the IR telescopes is larger in the plane perpendicular to the ecliptic, the difference from the volume in the ecliptic plane is small.

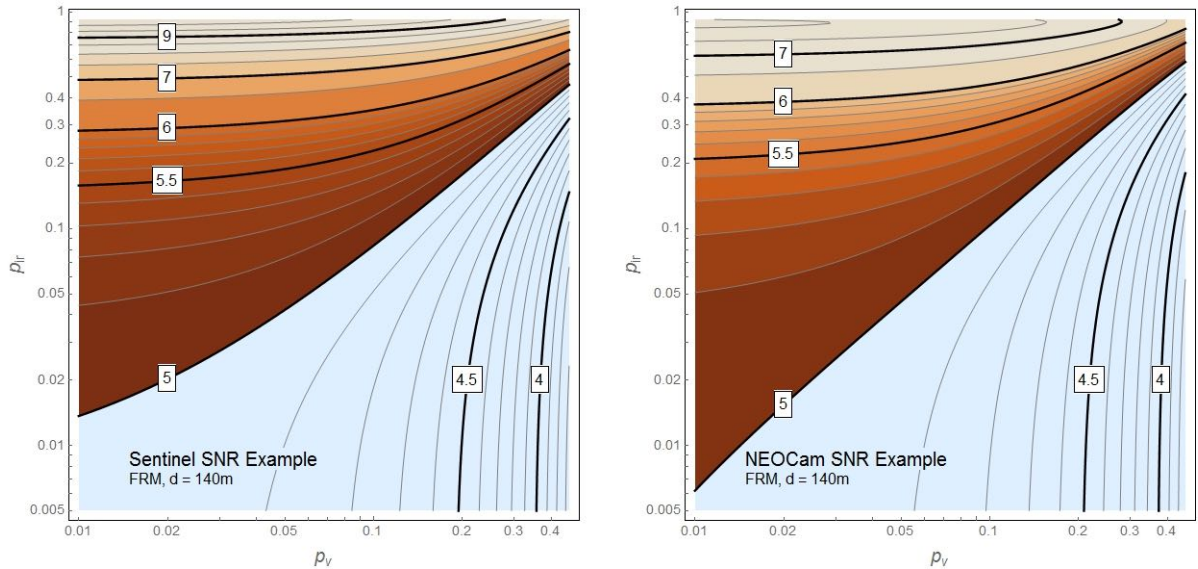

**Fig. 11.** Impact of $p_v$ and $p_{ir}/p_v$ on SNR (contour lines) for Sentinel (left) and NEOCam (right), given an asteroid of 140 m diameter position near the edge of the search volume for each instrument. Areas in parameter space where SNR ≥ 5 are shaded brown; those with SNR < 5 are in light blue. The contour lines have constant SNR.

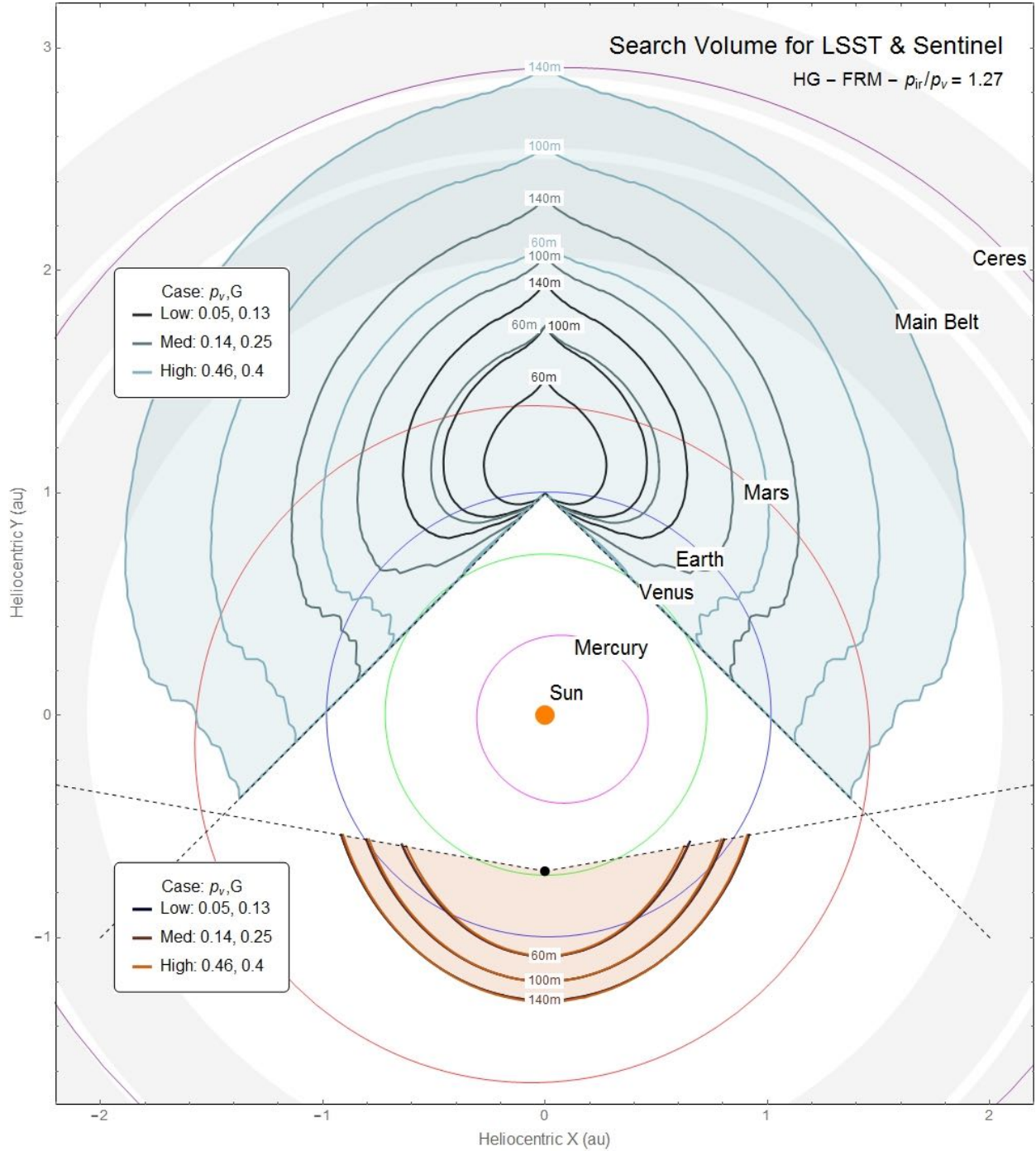

**Fig. 12.** Search volumes of LSST (blue region) and Sentinel (brown region) in the plane of the ecliptic for asteroids of diameter 60 m, 100 m, 140 m and for three scenarios of $p_v$ and G. Note that for 60m asteroids, the $p_v = 0.14$ boundary overlaps with those for 100 m and $p_v = 0.05$. Dashed lines show FOR boundaries. Volumes were calculated using the HG phase function, the FRM asteroid thermal model, and $p_{ir} = 1.27\ p_v$.

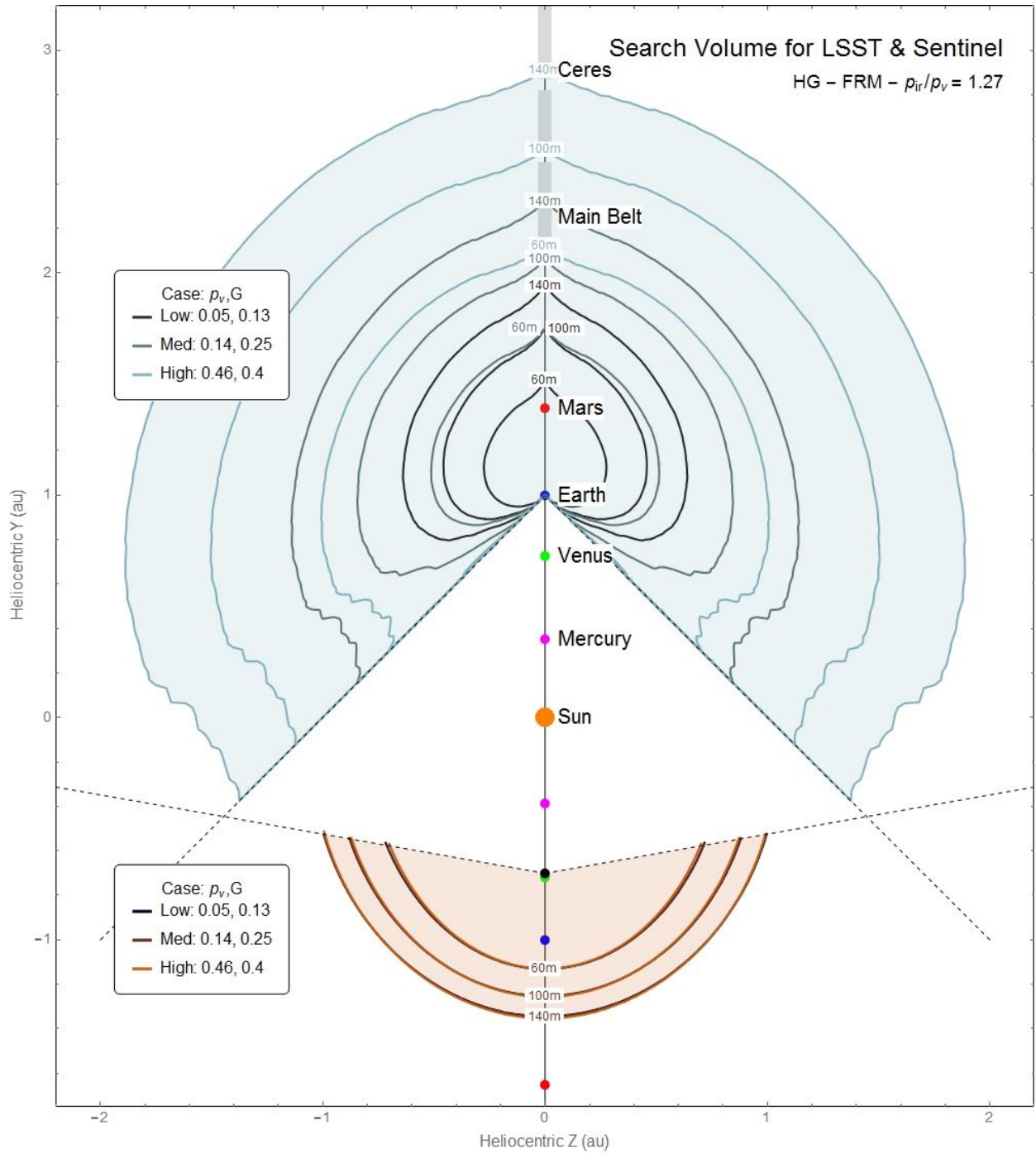

**Fig. 13.** Search volumes as in Fig. 12, but in a plane perpendicular to the plane of the ecliptic. The Sentinel search volume is slightly larger in this plane than in the plane of the ecliptic (*cf.* Fig. 12) because the noise floor $M_{\text{noise}}$ is lower in higher ecliptic latitudes.

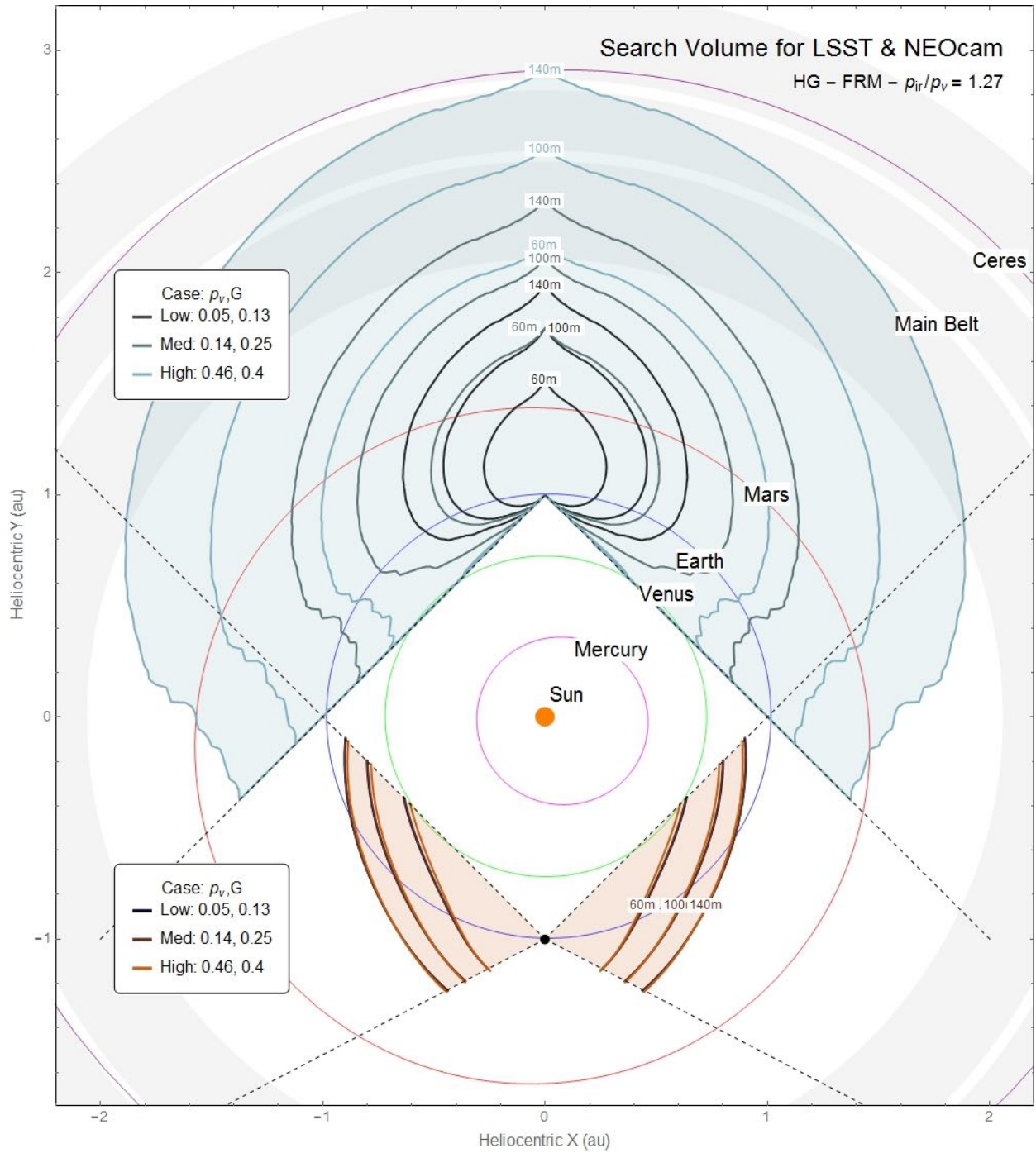

**Fig. 14.** Search volumes as in Fig. 12 for LSST (blue region) and NEOCam (brown region).

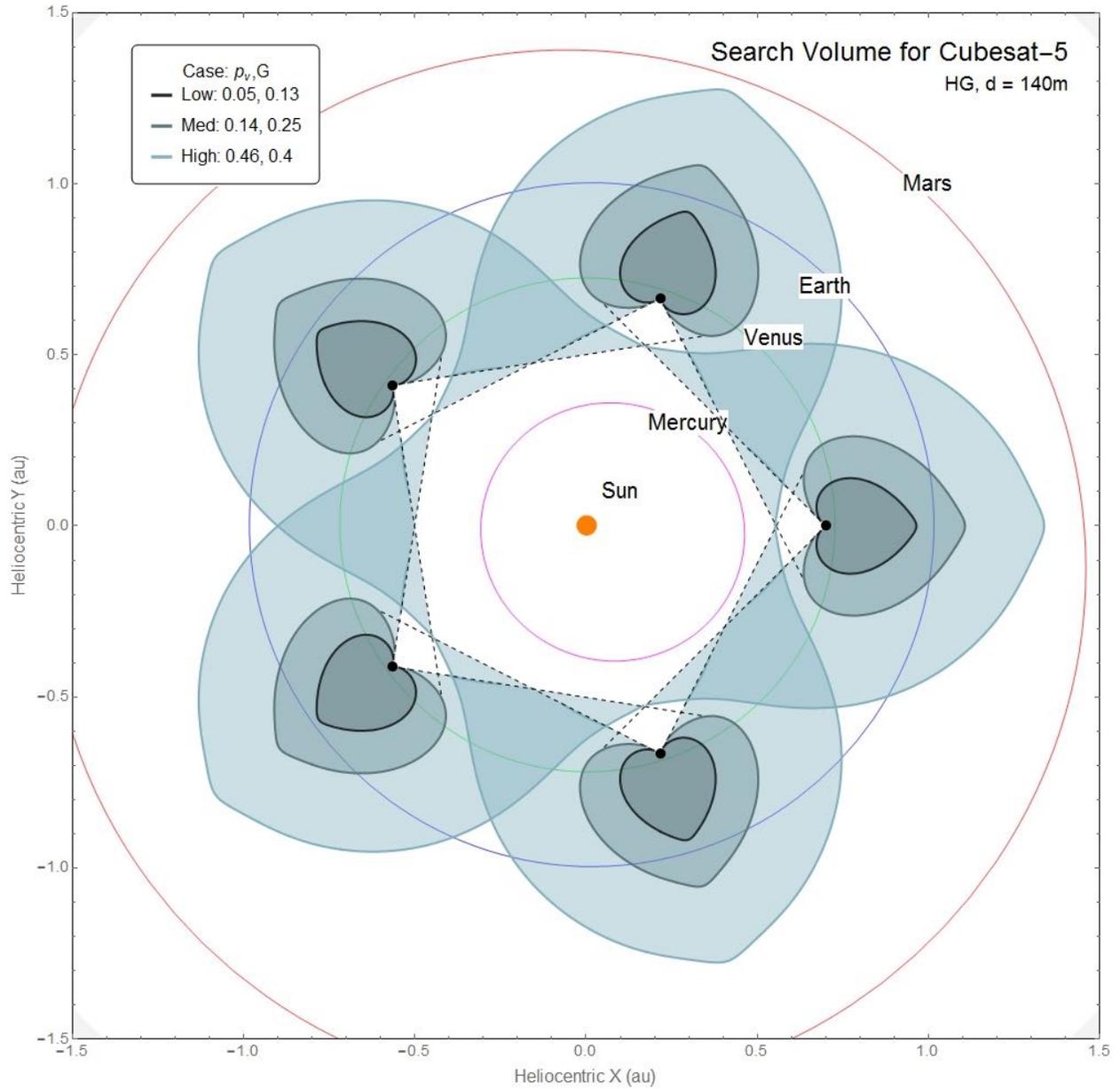

**Fig. 15.** Search volumes for the Cubesat-5 constellation in the plane of the ecliptic for asteroids of diameter 140 m and for three scenarios of $p_v$ and $G$. Dashed lines indicate FOR boundaries.

## 3.3. Visibility Ratios

The visibility ratio as a function of radius $r$ is defined by Eqs. (41) and (43) and Fig. 8. Fig. 16 and 17 plots these ratios for various asteroid diameters and albedo scenarios for the visual-band and IR-band telescopes, respectively. A visibility ratio equal to 1 signifies that all asteroids at radius $r$ would be visible to the telescope continuously. Ratios less than unity reflect the probability of visibility for asteroids at orbits defined by randomly distributed orbital elements $\Omega$ and $\omega$ at radius $r$.

Because all NEOs, by definition, have a perihelion $ph \leq 1.3$ au and an aphelion $ah > 0.983$, a "breadth-first" search strategy could constantly watch the annular region $0.983 \leq r \leq 1.3$ au with visibility ratio as close as possible to 1. Eventually all NEO would cross into this region for detection. Such a strategy, sometimes called a "wide survey" approach (Farnocchia et al., 2012; Hills and Leonard, 1995) searches a narrow depth in $r$ but has a high probability of detecting the every asteroid within that range. In contrast, a "depth-first search" or "deep survey"(Farnocchia et al., 2012; Morrison, 1992) would observe more deeply (in radial distance $r$, but that is closely related to apparent magnitude and diameter), while allowing for a low visibility ratio at each radius $r$.

Any real telescope will perform somewhere in between these two extremes, with a tradeoff that we can visualize by plotting the visibility ratio as a function of $r$ (Figs. 16 and 17). Single telescope systems (including LSST, Sentinel, and NEOCam) must look away from the Sun, and this limits their maximum breadth—a large region of space will always remain outside their FOR. As an example, consider asteroids of 140 m diameter and medium brightness ($p_v = 0.14, G = 0.25$) crossing $r = 1.14$ au. LSST achieves a peak visibility ratio of 0.54 in this case. Although the distance is fairly deep, the instrument would miss around 46% of the asteroids crossing at that radius because they are obscured by the Sun.

There are two ways that LSST could avoid missing those 46%. It could search a wide range of radii $0.7 \leq r \leq 2.4$. Some of the asteroids at $r = 2$ au, for example, will later be at $r = 1.14$ au. Alternatively, LSST could simply wait for the asteroids to move into the search volume. The wait might stretch to months or years, however.

The Cubesat-5 constellation could enjoy the highest visibility ratio because it would field multiple telescopes to survey different parts of the sky simultaneously. With enough satellites—or sufficiently bright asteroids—it might achieve the breadth-first search goal of keeping some annular region under constant surveillance (*i.e.*, a visibility ratio of 1.0), at least within the FOR and visibility approximation contemplated here. This is evident in Fig. 15, which shows that for the high case of bright asteroids ($p_v = 0.46, G = 0.4$) of $d = 140$ m, the search volumes of the five satellites overlap; as a result the visibility ratio for $d = 140$ m goes to 1.0 from $0.71 \leq r \leq 0.91$ au (lower right panel in Fig. 16). In each of the three albedo scenarios, the visibility ratio of the CubeSat-5 constellation exceeds that of LSST; the tradeoff is that its visibility ratios are high only in a narrow range of depths compared to those of LSST and Sentinel.

NEOCam would monitor the least overall depth in radius of any of the systems, although its peak visibility ratio slightly exceeds that of Sentinel. The nature of its FOR and the capture ratio of its sensor at $r \geq 1$ combine to give it a very narrow range in radius $0.98 \leq r \leq 1.33$. NEOCam would achieve a peak visibility ratio of 0.31 at $r = 0.99$. Sentinel's maximum visibility ratio of 0.33 would occur at $r = 1.07$ au, and its depth range would cover $0.69 \leq r \leq 1.35$.

### 3.4. Fraction of NEO within Search volume

Using the radial visibility ratios and the two radial distributions of asteroids shown in Fig. 2, I used Eq. (44) to calculate the fractions ($f_{ir}$ and $f_v$) of the NEO population falling within the search volume of each telescope at any moment in time, for both visible-band and infrared observations. Fig. 18 plots the results for the three albedo scenarios.

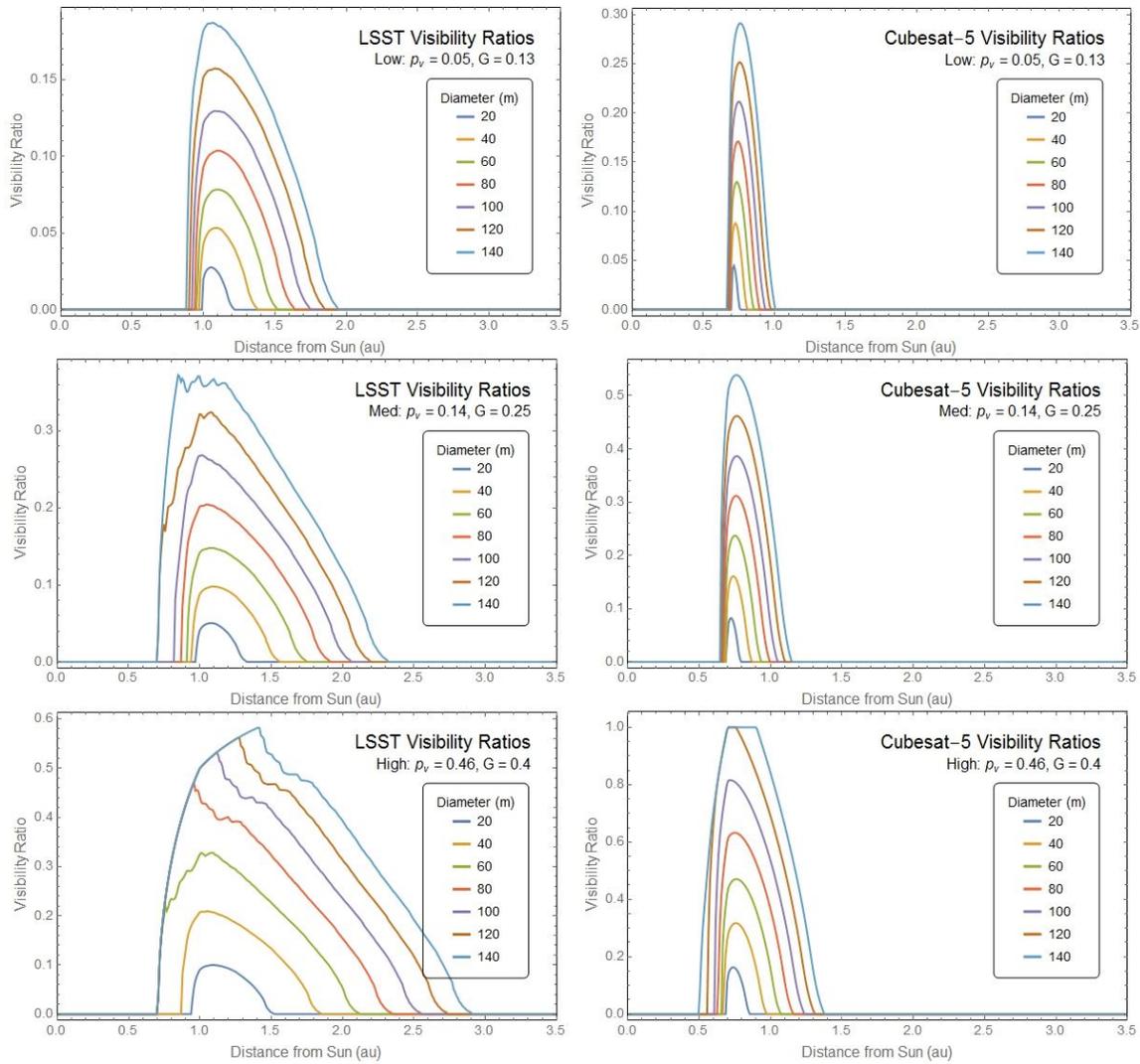

**Fig. 16.** Radial visibility ratios for the visible-light telescopes LSST and Cubesat-5 for three albedo scenarios and asteroid diameters up to 140m. Note that the vertical scale is different for each plot, but the horizontal scale is the same.

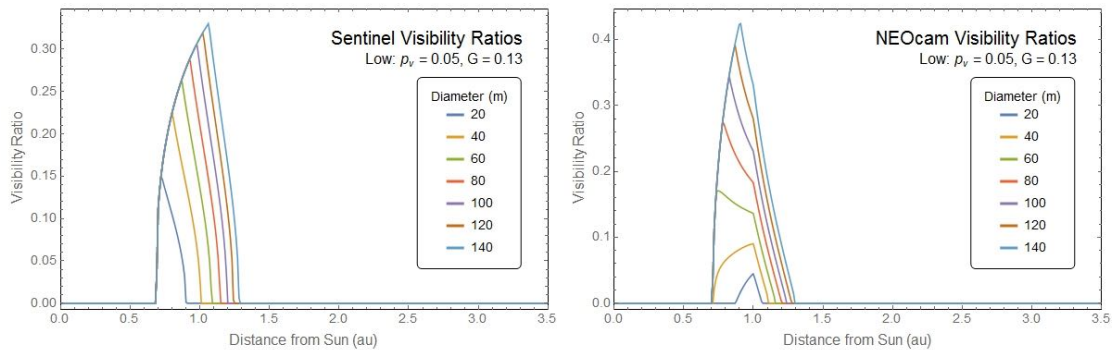

**Fig. 17.** Radial visibility ratios for the IR telescopes Sentinel and NEOCam for various asteroid sizes. Results for the Low albedo scenario are shown; results for the other two scenarios are very similar.

In the case of the general NEO distribution and low-brightness asteroids ($p_v = 0.05, G = 0.13$), Sentinel has the highest fraction of NEOs of size $20 \leq d \leq 102$ m in its search volume. For large asteroids of $102 \leq d \leq 1000$, LSST achieves the highest fraction. The Cubesat-5 constellation has the lowest fraction for NEOs of size $20 \leq d \leq 378$ m.

In the case of medium brightness asteroids ($p_v = 0.14, G = 0.25$), LSST has the highest fraction of asteroids, regardless of size, in its search volume. Sentinel is second best in the size range $20 \leq d \leq 135$ m, and Cubesat-5 is second best from $135 \leq d \leq 1000$ m. Cubesat-5 has the lowest fraction from $20 \leq d \leq 135$ m.

I find that the detection of bright asteroids ($p_v = 0.46, G = 0.4$) is likely to be dominated by the visible light telescopes LSST and Cubesat-5. For these asteroids, LSST has the highest fraction in a wide size range, $20 \leq d \leq 880$ m, while Cubesat-5 is second highest for nearly all sizes, followed by Sentinel. NEOCam has the lowest fraction in this albedo scenario.

The results for the impactor distribution are qualitatively similar, albeit with somewhat different crossover points (Fig. 18). In general, there is less advantage in performing a deep search for an impactor distribution that spans a relatively narrow radial range (Fig. 2).

Fig. 19 illustrates the relative performance on this metric of NEOCam vs. each of the other telescopes. I selected NEOCam as a point of comparison because its curves are relatively consistent, and it often has the lowest fraction. In the case of a general NEO distribution of high brightness asteroids, my results suggest that LSST will collect 12 times as many asteroids in its search volume at $d = 20$ m as NEOCam would, and 7 times as many at $d = 140$ m. This advantage drops to a factor of four at $d = 1000$ m. In the Low albedo scenario, the LSST response is never below that of NEOCam, and it ranges up to three times as high. The picture is qualitatively similar for a distribution of impactors, but the ratios are smaller.

## 3.5. Initial Rates of Detection

Using Eq. (46), I computed the total fraction ($\Gamma_v(p_v, G)$ and $\Gamma_{ir}(p_v, G)$) of asteroids of size $140 \text{ m} \leq d \leq 1\,000$ m that would fall into the search volume of each telescope as a function of absolute magnitude $Q(H)$. As shown in Fig. 20, LSST would have 16% of all low-brightness asteroids from the general NEO population that are in this size range in its search volume at any point in time, as well as 28% of medium-brightness objects and 41% of high-brightness asteroids. For the same distribution, NEOCam has projected detection of 7.5%, 7.4%, and 7.3% of low-, medium-, and high-brightness asteroids, respectively; Sentinel's numbers are nearly the same. Total fractions are generally higher for the impactor distribution (Fig. 20).

The statistical calculations here as elsewhere assume that the NEO population has a distribution that is approximately invariant with time. Asteroids will enter and exit the search volume, but the number visible at any one moment remains approximately the same.

The large differences in total fractions among the four telescopes highlights the consequences of their very different search strategies. NEOCam's breadth-first search is quite narrow in radius $r$. As a result, it sees a relatively small number of asteroids at any one time. Over time, asteroids will move through its search region, and its search region will move through space as well. Its list of discoveries would thus be expected to grow steadily. In contrast, LSST will conduct a search that is both wide and deep in $r$. As a result, it will have a huge number of asteroids in its search volume at any point in time.

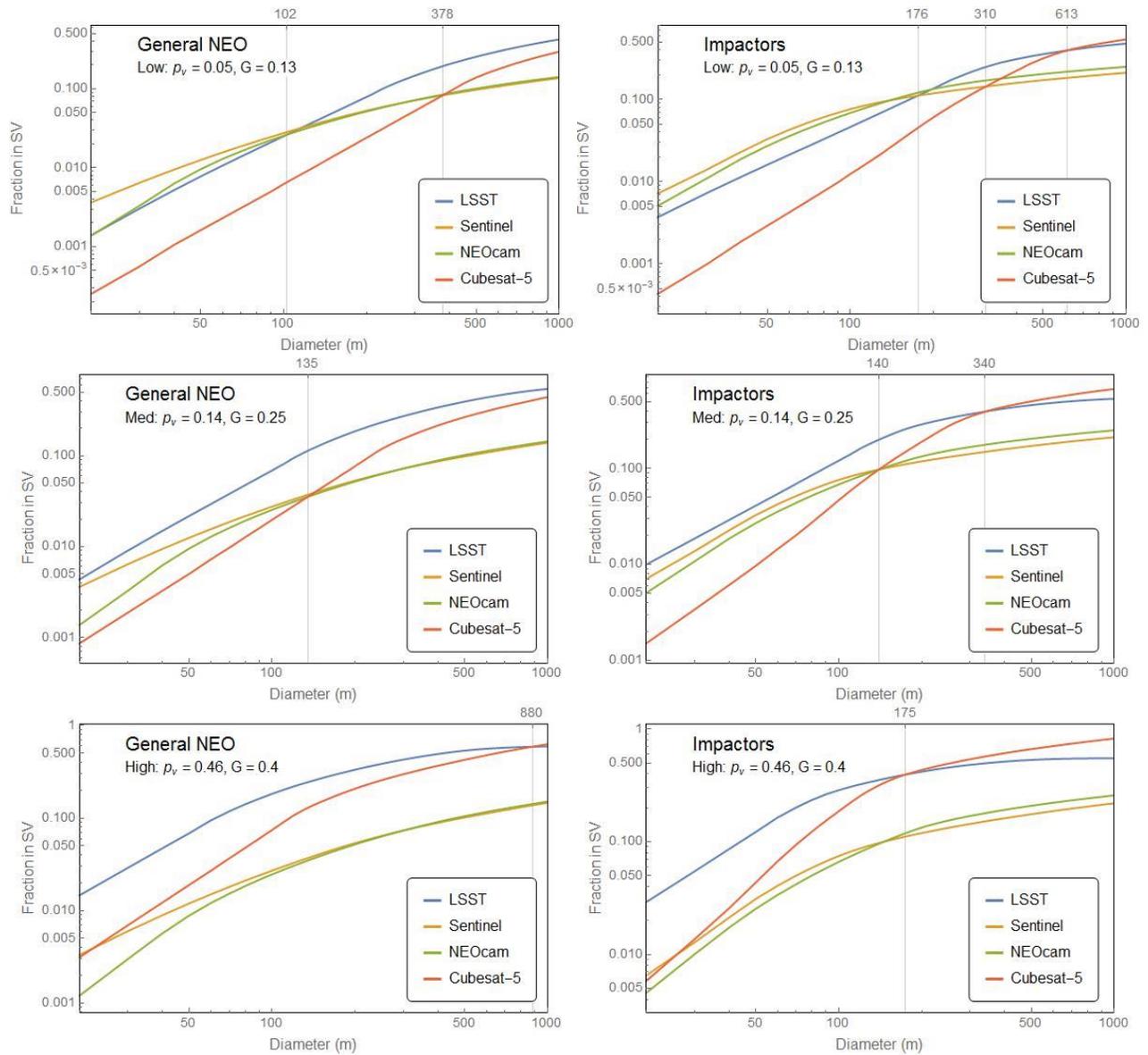

**Fig. 18.** Fraction of NEO populations within the search volume at any point in time. Each plot shows results for all four telescopes and for asteroids of size $20 \leq d \leq 1000$ m. Low, Med, and High asteroid brightness scenarios are shown. Asteroid diameters at which one telescope curve crosses another are marked with vertical lines.

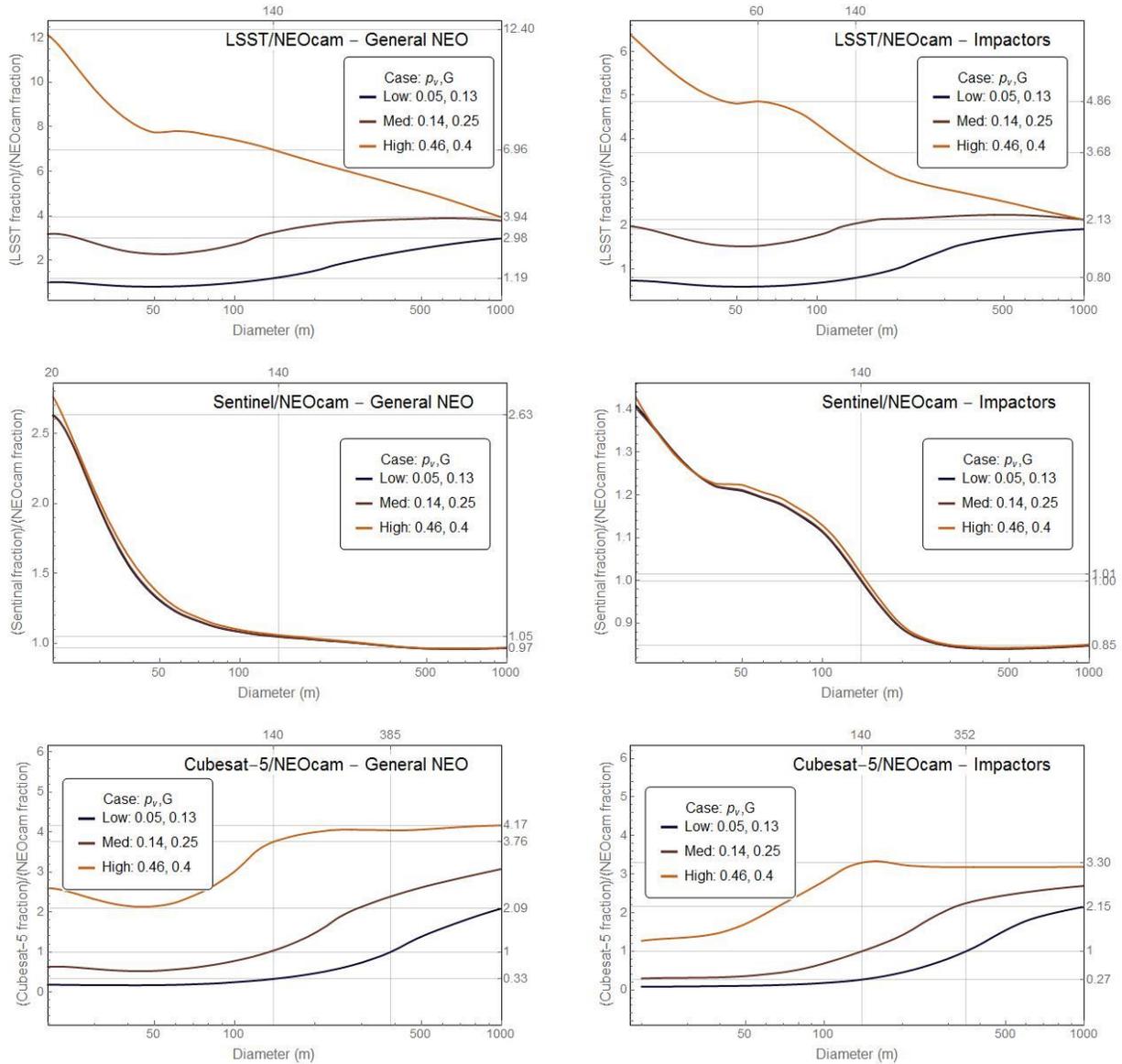

**Fig. 19.** Ratios of the NEO population fractions in the search volumes of LSST, Sentinel, and Cubest-5 to that of NEOCam, for asteroids of diameter $20 \leq d \leq 1000$ m. Vertical lines mark $d = 140$ m and diameters at which a maximum ratio is achieved or the ratio crosses 1.0.

At the commencement of the search, each telescope has the potential to detect and discover all of the asteroids that are then in its search volume. Considering only asteroids of size $d \geq 140$, the results here suggest that LSST has the potential to detect ~28% of all general NEOs and ~36% of all impactors within its first three-day cadence cycle. Sentinel has the potential to discover ~9% of general NEOs and ~16% of impactors in its first 24 days; NEOCam to find ~6% of general NEOs and

~12% of impactors in its first 11 days. Cubesat-5 could find 18% of general NEOs and 34% of impactors within its first cadence cycle. In addition to the GEB-relevant asteroids of $d \geq 140$ m, the telescopes would also discover many smaller NEOs within these initial cadence cycles.

Whether the instruments realize these detection potentials or not depends on the search efficiency of the observing cadence and other factors that are beyond the scope of this study. These results should be considered upper bounds that are unlikely to be fully realized in practice.

Notwithstanding the impressive initial rate of discoveries, completing the GEB (or any comparably high-percentage survey) will take years because the median orbital period of general NEOs is 3.1 years; 10% of them have a period greater than 4.8 years. Long-period NEOs may take years to come into view regardless of the telescopes search volume.

The disparity among telescopes in initial search depth has been unclear in most previously reported survey simulations, in which the primary result was an estimated total time to complete the GEB. Although the 90% completion time does depend somewhat on the telescope, the primary driver is orbital dynamics. The initial detection rate highlights telescope-specific differences more clearly.

After the first week of operation, the detection rate will drop for any instrument because the search volume is mostly the same as it was for the first week. Changes to the search volume occur for two reasons. The first is that NEO orbits will take them in and out of the search volume over time. The rate at which this occurs is difficult to predict with the simple methods used here and awaits more sophisticated simulation.

The second reason the search volume changes is that that the telescope is orbiting the Sun; as its position in space changes, so does its search volume. The left panel of Fig. 21 illustrates this effect by charting how Sentinel's search volume will shift from the start of one 24-day cadence cycle to the beginning of the next. The larger the search volume and the shorter the cadence cycle, the smaller the cadence-to-cadence change is as a fraction of the total. In the case of Sentinel, one cadence cycle changes the fraction of 140 m NEOs by 45%. At the other extreme, for LSST in the High albedo scenario, the fraction of 140 m NEOs changes by only 2.3% over the course of a three-day cadence cycle.

In the case of Cubesat-5, its low limiting magnitude results in a 100% cadence turnover for asteroids with $d \leq 140$ m (Fig. 21). The search volumes for two successive cycles thus have no overlap. This strongly suggests that Cubesat-5 will not be able to complete one pass through its search volume in its ~25 day cycle before the search volume shifts to another part of space.

I integrated the cadence turnover with the size distribution and divided by the cycle length to calculate the number of asteroids newly appearing in the search volume per day (bottom panels of Fig. 20). (Cubesat-5 was omitted because its successive search volumes do not overlap.)

During LSST's second 3-day cadence cycle, the majority (~90%) of the asteroids in its search volume will be the same as the previous cycle. Some of these asteroids will have moved out of the first search volume (*i.e.,* move from the red to the blue in the left panel of Fig. 21), but at least during the second 3-day cycle most will be not be new. As a rough approximation this gives us the initial detection rate after the first cycle.

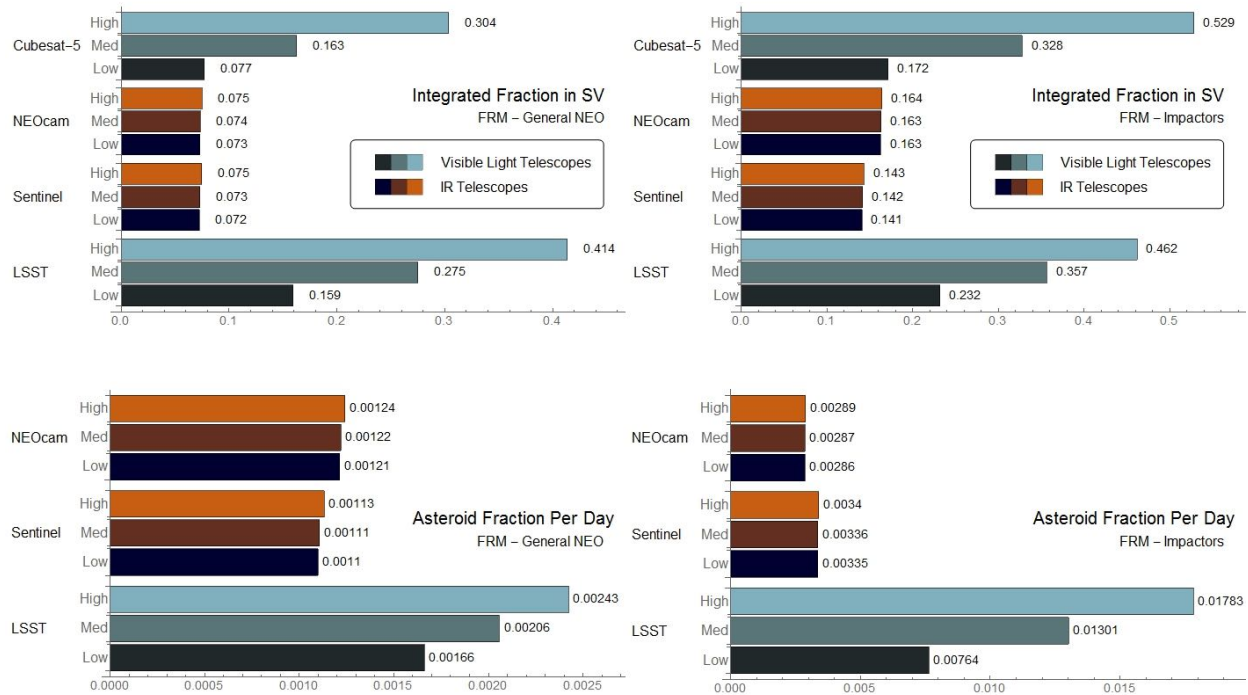

**Fig. 20.** Integrated NEO population fraction and by albedo class in search volume for each telescope (top panels) and the change per day in the fraction in the search volume due to search volume shifting (bottom panels). The plots show results for asteroids with $140 \leq d \leq 1\,000$ m.

In its second cycle, LSST will also get another chance to detect the 90+% of asteroids that were also present in the first cycle, some of which might be missed due to observing issues. Extended observation of already-discovered asteroids also allows higher-precision measurement of orbital parameters and the construction of light curves.

Calculating the detection rate over time is beyond the capability of the simple techniques employed here. As the percentage of the population which has been discovered rises, the new detection rate must fall.

An important finding here is that many benefits of the search will arrive well before the end of the survey. It is possible that the first third of the survey could be completed very quickly—literally in days or weeks—with some telescopes. This is an important consideration for space telescopes, whose service lifetimes are hard to predict. Early termination of a mission, as occurred with the Kepler space telescope, will yield partial results. Detection rate over time and interim results are relevant metrics for assessing the impact of early termination, and should be part of future simulation results.

## 3.6. Results for GFRM, NEATM, and Tumble Thermal Models

The results reported above reflect the use of FRM to model asteroid thermal characteristics. For comparison, I also used the alternative models NEATM, GFRM, and Tumble (Fig. 6) to compute visibility fractions (Fig. 22) and search volumes (Fig. 23).

As a generalization, it appears that when $\eta$ is allowed to vary within the range typically found in observational studies, the results for NEATM are qualitatively similar to those for FRM, but with a wider band of variation. Some NEATM and GFRM models overlap; this begs the question of whether some asteroids that are well fit in previous analysis by NEATM might be better fit by GFRM. Further research is needed to address that question.

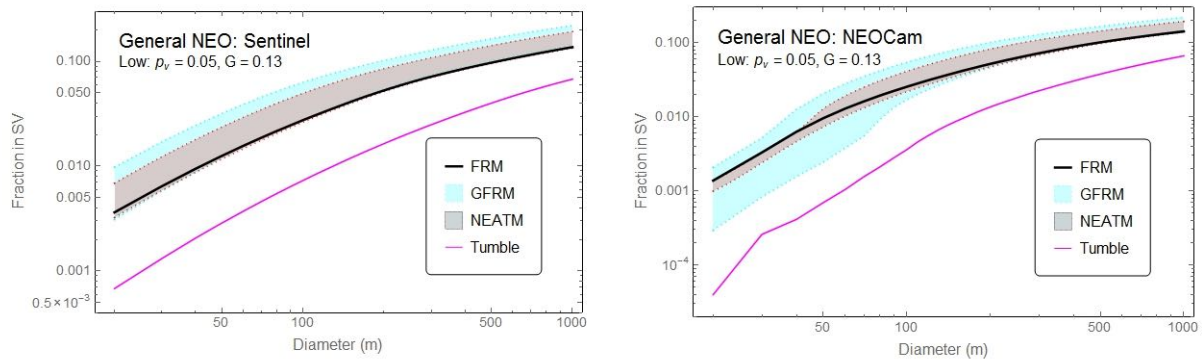

**Fig. 22.** Comparison of visibility fraction results for four thermal models for the IR telescopes Sentinel (left) and NEOCam (right), using the general NEO distribution. Results for FRM, shown here as a thick black line, are presented in more detail in Fig 17. The envelope of NEATM results for $0.66 \leq \eta \leq 2.24$ (gray region) is wider for Sentinel than for NEOCam. In contrast, the envelope of the GFRM cases depicted in Fig. 6 (light blue region) is broader for NEOCam at asteroid diameters of 100 m or less. The magenta curves plot results for a fast tumbling asteroid having high thermal inertia. The behavior for the impactor distribution (not shown), and for the Med and High scenarios are similar.

When using the Tumble model, which applies the reasoning used for FRM to the case of a tumbling asteroid, search volumes and visibility fractions roughly parallel those found using FRM but are much smaller. The signal from an asteroid modeled with Tumble with parameters $d, G, p_v, p_{ir}$ is roughly equivalent to an asteroid using the FRM model with the same values of $G, p_v, p_{ir}$, but with a much smaller diameter $d$. In the Low albedo case (*i.e.*, $G = 0.13, p_v = 0.05, p_{ir} = 0.0635$), an asteroid with $d = 140$ m under Tumble gives the same signal as an FRM asteroid with $d = 45$ m— and, conversely, to achieve under Tumble the same signal as FRM gives at $d = 140$ m requires increasing the asteroid size to $d = 478$ m. Tumbling asteroids thus appear to be very difficult to find by using an IR telescope, as discussed further in section 7.

The Tumble model used here assumes the worst case of fast tumbling. An asteroid having fast rotation in a principle axis that slowly precesses or wobbles would presumably be more accurately modeled by GFRM.

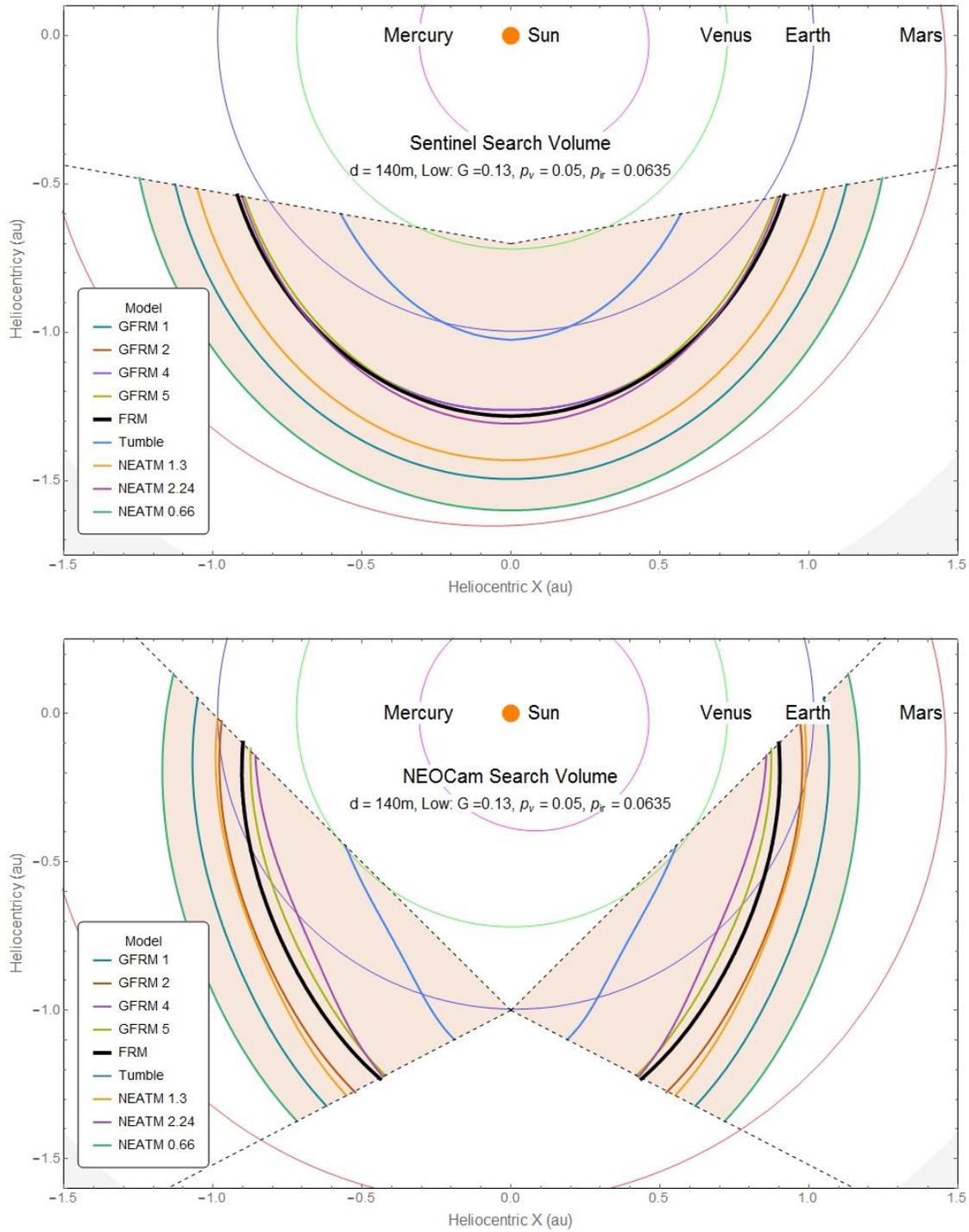

**Fig. 23.** Comparison of search volumes for Sentinel (top) and NEOCam (bottom), calculated using FRM, GFRM, NEATM, and Tumble thermal models. The GFRM cases are depicted in Fig. 6. NEATM is shown for three values of the fitting parameter $\eta$. Because NEATM and some GFRM cases have strong phase effects, the curves can intersect. FRM, GFRM 4, GFRM 5 and NEATM $\eta = 2.24$ in particular are very similar for both telescopes. The curves for NEATM with $\eta = 1.3$ and for GFRM 2 overlap for NEOCam, whereas for Sentinel GFRM 2 clusters with FRM.

## 4. Sensitivity Analysis

The NEO search problem is difficult in part because so many of the pertinent variables and parameters can vary over a wide range. To establish which of these factors matter most, I performed sensitivity analyses to estimate or bound the effects of certain assumptions.

### 4.1. Sensitivity to Asteroid Parameters

Throughout this study, I use three albedo scenarios (Low, Med, and High) and associated values of $G, p_v,$ and $p_{ir}$. Fig. 24 examines how well these three scenarios capture the range of variation in an asteroid population observed by Sentinel and NEOCam. Three cases are plotted against a range of values obtained by taking a data set of 583 asteroids for which $G$ and $p_v$ have been measured (Pravec et al., 2012) and found to fall within the ranges $0.025 \leq p_v \leq 0.472, -0.14 \leq G \leq 0.51$. The IR albedo $p_{ir}$ was added at both $p_{ir} = 2\ p_v$, and $p_{ir} = 0.5\ p_v$. The search volume for each case was plotted; the shaded red regions in Fig. 24 indicate the envelopes that contain all of these curves.

Fig. 24 demonstrates that the three test cases essentially fall in the middle of the ranges obtained from this population of 583 asteroids. For both telescopes, the case that yields the widest region is $G = 0.24, p_v = 0.419, p_{ir} = 0.837$, and the case that forms the inner boundary is $G = 0.46, p_v = 0.476, p_{ir} = 0.238$. Interestingly, both the upper and lower bounding cases have relatively high values of $p_v$. In addition, these two parameter sets do not generate either the highest or the lowest value of $T_{FRM}$ among the points in the range tested. These results reinforce the important roles that the IR albedo $p_{ir}$ and IR reflection play in determining the search volume.

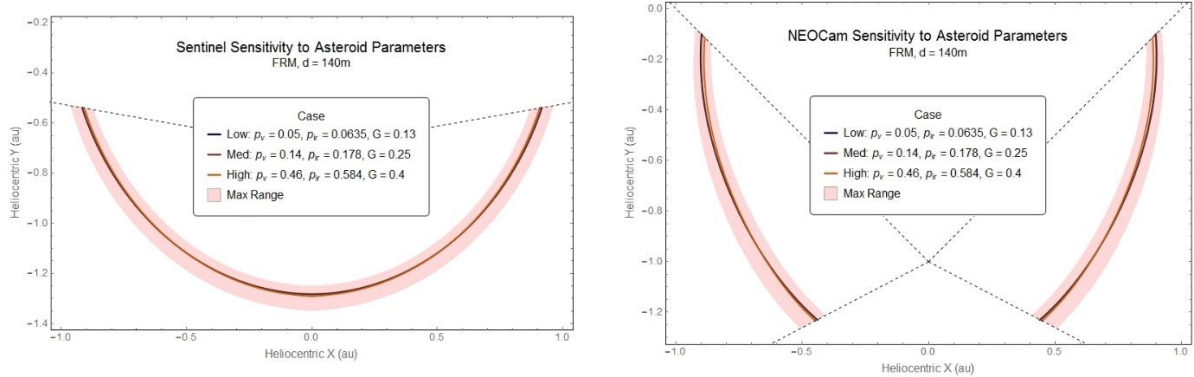

**Fig. 24.** Comparison of search volumes obtained from test cases to those produced by diverse asteroid parameters for Sentinel (left) and NEOCam (right). The Low, Med, and High scenarios (Fig. 12 and 14) for asteroid diameter $d = 140$m are drawn as solid lines. Results from the parameter ranges found empirically in a data set of 583 asteroids are shown as shaded red regions.

## 4.2. Sensitivity to IR Noise Threshold

The noise floor $M_{\text{noise}}$ is important to the IR telescopes. The noise floor varies with the zodiacal dust density, as well as with direction and season. Fig. 25 shows the impacts of reducing the noise threshold to 75% of its nominal values, and of raising it to 150%. The resulting variations are quite modest; they are smaller than the variations obtained from the sensitivity analysis for asteroid parameters (Fig. 24).

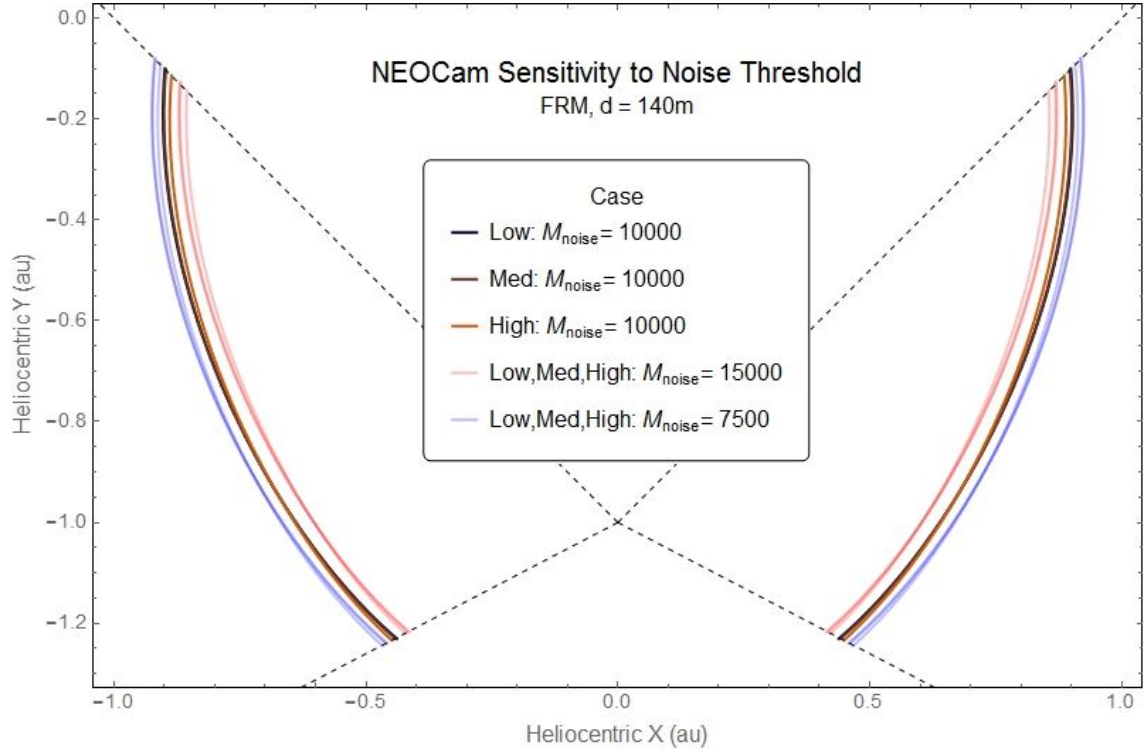

**Fig. 25.** Comparison of search volumes for test cases (noise floor of 10,000) to those resulting when the NEOCam noise floor was lowered to 7500 or raised to 15,000, for asteroids of diameter $d = 140$ m and three albedo scenarios. Results for Sentinel (not shown) were very similar.

## 4.3. Sensitivity to LSST Limiting Magnitude

All LSST calculations in this study have used the median case of the statistical distribution of limiting magnitude with solar elongation angle from Fig. 7. Fig. 26 shows the effect of using other choices. The median case is very close to a constant $V_{\text{limit}} = 24.6$.

## 4.4. Sensitivity to SNR Threshold for Detection

A critical measure of any survey is the $SNR$ threshold used for detecting asteroids. Typically this is set so that $SNR \geq 5$ to provide a reasonable balance between sensitivity and the probability of errors. The threshold value reflects an intuition that many of the small, faint, or distant objects that the survey detects will be observed at $SNR \approx 5$.

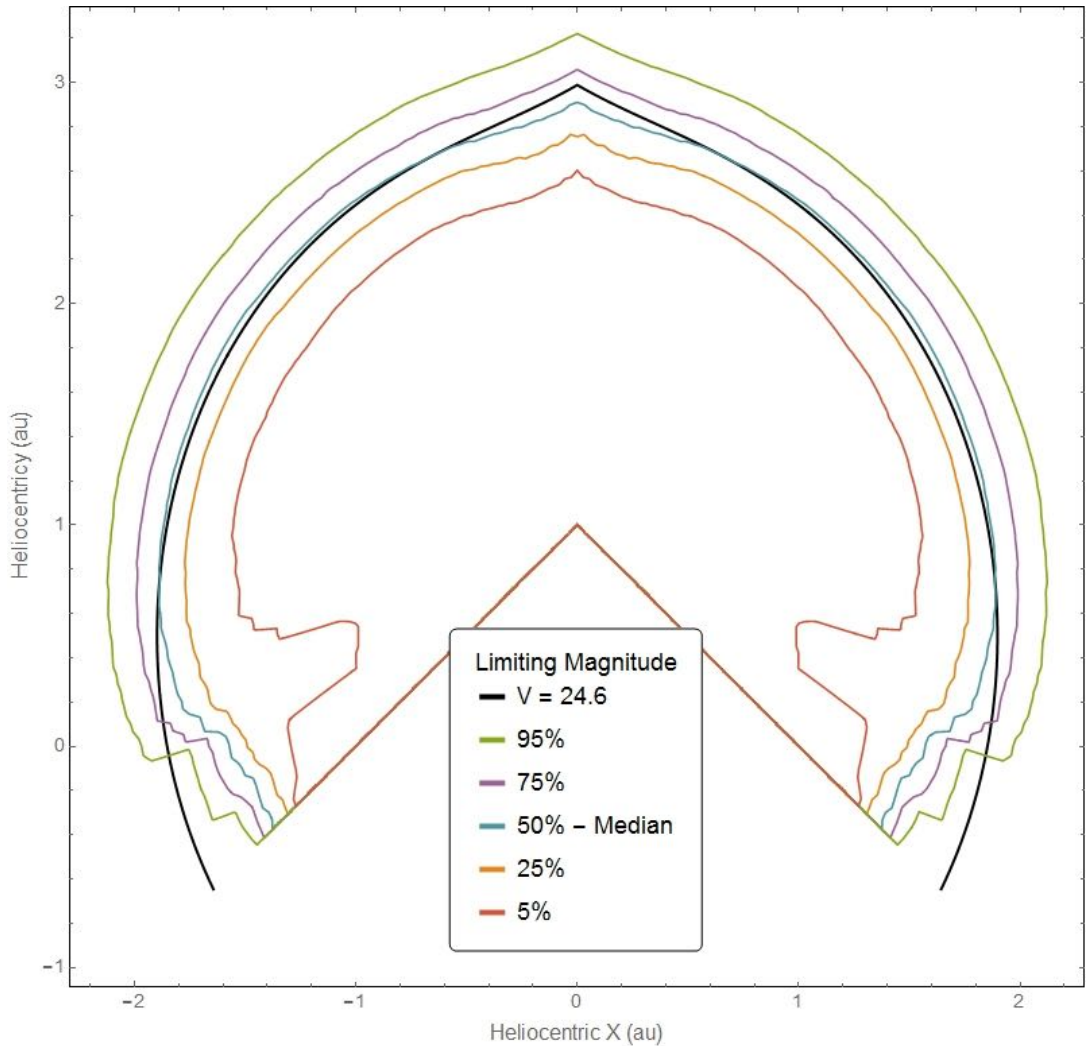

**Fig. 26.** Search volumes for LSST calculated using a selection of limiting magnitudes from Fig. 7. The idealized version (limiting magnitude of 24.6, black) does not account for atmospheric or terrestrial effects. At the median magnitude, the search volume is very similar in size to that of the idealized case except for deviations at low solar elongations and, to a lesser extent, at zenith. The search volumes for 75% and 25% magnitude values closely bracket that of the median.

I find, surprisingly, that this is not the case. The intensity of sunlight diminishes en route to the asteroid in accordance with the inverse square law, and light reflected or transmitted from the asteroid also falls off en route to the telescope as the inverse square of the distance. The combination of these two effects imposes a high degree of nonlinearity on the observations. IR telescopes experience a further nonlinearity because their sensor passbands admit only a fraction of total thermal emissions. Small differences in temperature shift the Planck spectra of the asteroids in or out of these passbands. As a result, IR observations are very sensitive to distance.

As seen in the top panel of Fig. 27, the SNR climbs very steeply inside the search volume. The lower two panels of Fig. 27 shows how coverage of asteroids within the search volume (expressed as a fraction of the total seen at $SNR \geq 5$, for the general NEO population) declines as the minimum detection $SNR$ increases. The falloff is much less steep for the IR telescopes than it is for LSST. Surprisingly, at an SNR threshold of 6, Sentinel would be expected to detect 89.4% of the asteroids of size $d = 20$ m that it could detect at an $SNR \geq 5$; for asteroids of size $d = 1000$ m, that detection fraction rises to 96.5%. Similarly, I find that NEOCam would achieve $SNR \geq 6$ detection of 82.6% of those asteroids in its search volume detectable at $SNR \geq 5$ for $d = 20$ m, and 96.0% for $d = 1000$ m.

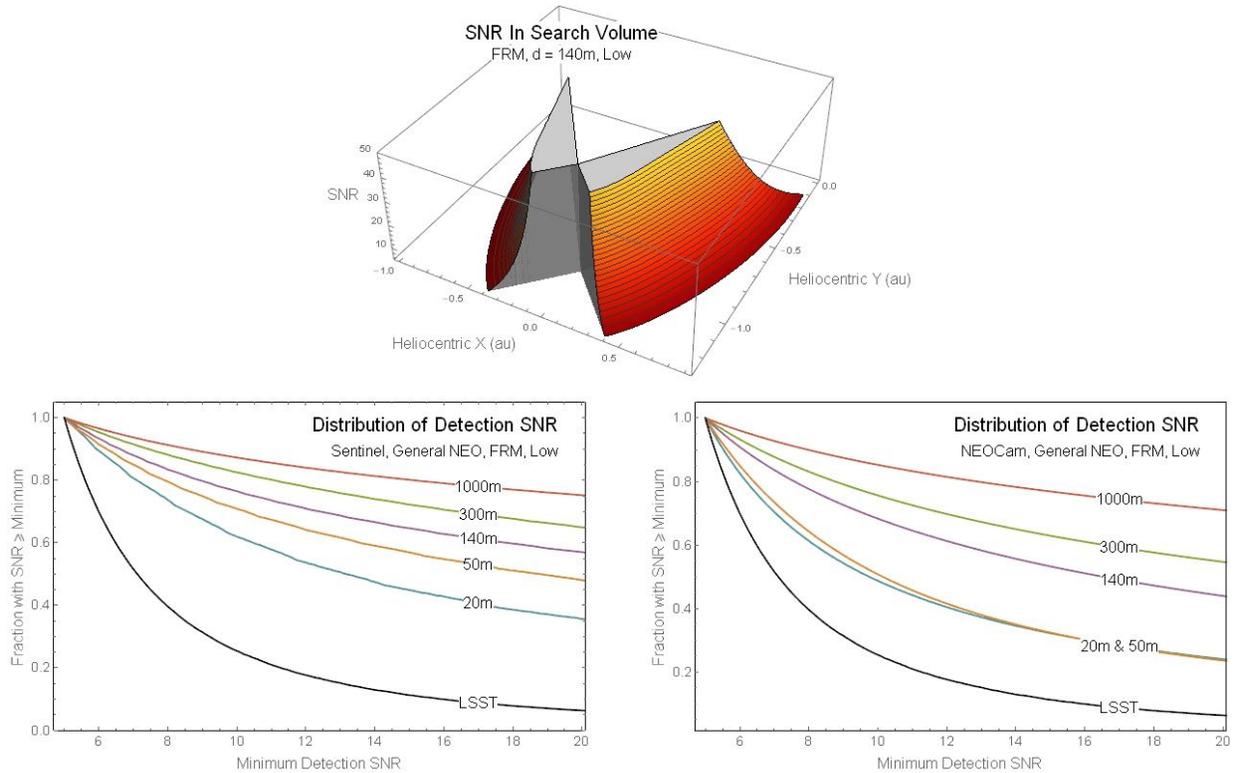

**Fig. 27.** $SNR$ for NEOCam plotted on top of its search volume (top panel) for an asteroid of $d = 140$ m and the Low albedo scenario. The bottom contour is $SNR = 5$, which defines the edge of the search volume. Even though the search volumes are defined by $SNR = 5$, the majority of asteroids are actually detected at a higher threshold. In the case of the IR telescopes this is asteroid size dependent, for LSST is it is not. Cubesat-5 results are not shown.

These high fractions should not be misunderstood as suggesting that $d = 20$ m asteroids are easy to find. Because the search volumes for small asteroids are very small, the likelihood of detection at $SNR \geq 5$ is not high; it is much lower than for large asteroids of $d = 1000$ m. But it is interesting that when a telescope—and in particular an IR instrument—is able to detect them, the detection is likely to be at a higher $SNR$ than the threshold.

One implication of this result is that we should expect only a small fraction of observations to have $SNR \approx 5$. Although the fraction of observations will generally differ from the fraction of the search volume, the two are closely related. In order for an asteroid of $d = 20$ m to be detected by NEOCam at $5 \leq SNR \leq 6$, all of the multiple observations of the asteroid must occur within a thin ring around the edges of the search volume that comprises just 17.4% of the volume. The interplay of the asteroid orbit and the telescope cadence makes calculation of how likely that is beyond the simple techniques used here.

One reason that LSST's $SNR$ function (Ivezić and Council, 2011) falls off less steeply than does that of the either IR telescopes is that reflected solar illumination lacks a sharp peak that moves in wavelength as a function of asteroid distance from the Sun. In addition, LSST's exposure time is not optimized for maximum $SNR$ but rather for others factors, such as rapid coverage of the $FOR$. A longer exposure time would improve $SNR$ by $\sqrt{t}$ and would also expand the search volume slightly, but it would also add more $SNR$ to the interior of the search volume. Given the current exposure parameters, 70.1% of the LSST search volume falls within $SNR \geq 6$, which seems quite adequate for its purposes.

## 5.     Discussion

This study has its origins in discussions with colleagues in the NEO search community that revealed a diversity of opinion on even basic assumptions. As one example, space-telescope advocates expressed extreme skepticism that a terrestrial telescope could search for NEOs at solar elongations $\gamma < 90°$ (e.g., Fig. 3.5 of the NRC Report (Shapiro et al., 2010), which directly inspired my Figs. 12–15). Meanwhile observational astronomers assured me that it was possible. This contradiction seemed worth resolving.

The analyses here suggest that while it is indeed more difficult to see an object low to the horizon and near the Sun, due to terrestrial effects summarized in section 2.7, detection under such conditions is far from impossible—low elongation simply changes the limiting magnitude and makes cadence planning more difficult.

A second example is the phase law effect: the fact that reflected light received from an object dims with as the object's phase angle increases. My analysis finds that this effect does not cripple observations using reflected light, as is sometimes suggested. These and the other issues discussed above are questions of quantitative limitations, and the goal of this study is to estimate their impact.

Detailed simulations that apply these principles to a possible NEO search in a detailed Monte Carlo simulation would allow greater detail, but it is hard to build such a simulation without the right base assumptions. I have shown here that some of the assumptions used by different groups are incompatible.

### 5.1.     Omissions and Directions for Future Work

The results presented in section 3 explore the instantaneous search volume of each telescope. While this snapshot provides insights, it is worth considering how it relates to the results of a multi-year survey. Simple visibility, as calculated here, places an upper bound on the performance of the telescope—an instrument clearly cannot detect asteroids that are not visible to it. In practice, however, additional factors will limit actual discoveries to a rate below this upper bound. The results here are not fully indicative of the actual performance, but it might be possible to simulate that using a much more detailed modeling approach than the one presented here. Notably, such a simulation would address the following issues that were omitted from consideration in the present study.

5.1.1. Cadence Sampling of Search Volume

The most serious omission from this analysis is that $FOV \ll FOR$, so the telescope must make an incomplete sample of the $FOR$. As a result, it is probable that a given cadence will miss some asteroids if, for example, they move outside the search volume before they can be adequately observed.

All observing cadences in NEO searches visit the same portion of the sky two to four times separated by 0.25 to 1.5 h. These visits allow the motion of the asteroid to put it in a different position for tracklet tracing. After the closely spaced repeat visits, the telescope does not revisit that portion of the sky until it has completed a pass through the $FOR$.

With the exception of LSST, single-pass durations through the FOR vary from 11 days for NEOCam to 24 days for Sentinel and 25 days for Cubesat-5. These periods are lengthy enough that one should assume that most asteroids seen in the multiple visits during the first pass will no longer be visible when the telescope next samples that portion of the sky. For this reason, these telescopes plan to use three to four visits per pass.

Kubica et al., 2007 conclude from their simulations that a cadence of two visits per night and a three-day return time would be feasibly and sufficient for LSST, and this has been the plan of record for that telescope (Ivezić et al., 2006). Other teams have expressed a great deal of skepticism to me that this will be possible. The skepticism appears to arise for two primary reasons.

The first is an apparent misunderstanding among some in the community that LSST will record as detected an asteroid seen on only two visits. In fact, the LSST plan (Ivezić et al., 2008b) is to require four or more observations to report a detection; in general, no more than two of those observations will occur on the same night. A typical detection would involve two observations on a single night followed by two more observations three days later. To succeed with this approach, enough asteroids must remain within the search volume for 3+ days to be spotted twice. Moreover, the LSST system will need to store potentially millions of tracklets from incomplete observations and then, three (or potentially more) days later, winnow them to accurately match tracklets to objects. The feasibility of these propositions were tested in simulations and have been reported to work successfully (Ivezić et al., 2006; Kubica et al., 2007). Whether this works in practice is open to question but at the very least objections leveled against this plan should reflect the actual propositions made by the LSST team (i.e. 4 observations, not 2).

A second source of skepticism about the LSST cadence plan is a misperception that the Pan-STARRS1 NEO survey attempted to use a two-visit cadence and was plagued by enormous number of false-positive detections. In fact, Pan-STARRS1 was plagued with systematic image artifacts at a rate orders of magnitude higher than expected. As a result of this and a sensor fill factor that was lower than expected, Pan-STARRS1 did not implement the two visit per day protocol (Denneau et al., 2013). The abortive Pan-STARRS1 experience thus did not clean test the feasibility of a cadence of two visits repeated after three days.

While the doubts expressed recently about the LSST cadence seem to be based, at least in part, on factual misunderstanding, this is no guarantee that LSST will be successful. As the Pan-STARRS1 experience demonstrates, the LSST team will need to be extremely careful about systemic errors and will have to demonstrate that their approach is robust to some achievable rate of both systemic and statistical errors. This is the topic of ongoing work by the LSST team.

More generally, the cadence-cycle turnover shown in Fig. 21 provides a useful metric of how much the search volume changes during each cadence cycle. The higher the turnover, the harder it is for the telescope to keep up. In the case of Cubesat-5, its cadence poses a severe turnover challenge to detection of all but very large asteroids because the search volume from one pass to the next is entirely new. Sentinel will see a 45% new volume each cadence cycle, NEOCam 25%, and LSST 2.3% to 7%, depending on the brightness.

A cadence-related issue for LSST is its limited observing time for regions at low solar elongation (*e.g.*, $45° \leq \gamma \leq 60°$). As shown in section 2.7, it is possible for LSST to observe those regions of space from the Earth, although it will have to deal with higher air mass and other terrestrial issues. But in order to succeed in these observations, LSST will have to adopt a special cadence to search in the best areas of the sky (*i.e.,* near the ecliptic) in the very short time period after sunset or before sunrise.

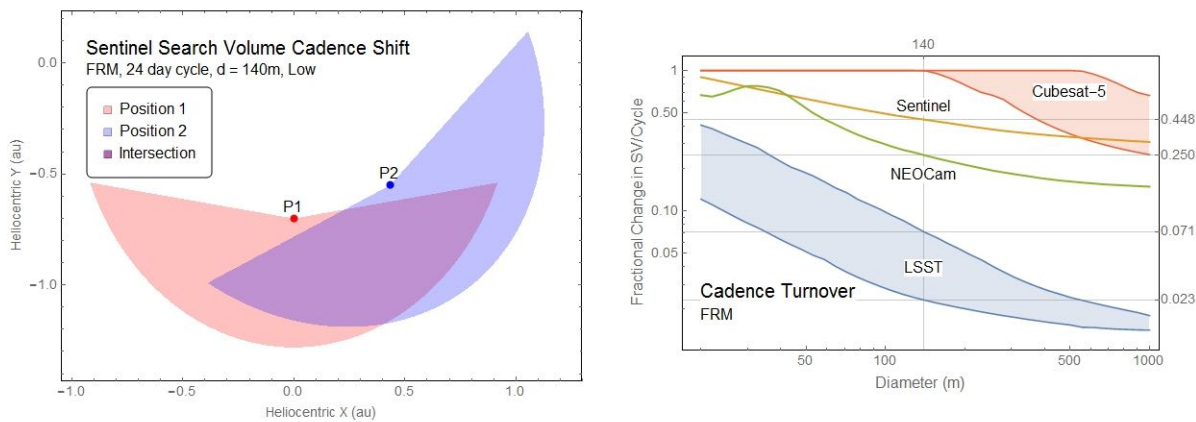

**Fig. 21.** The shift in the visibility during a cadence cycle (left), using Sentinel as an example. At the start of its 24-day cadence cycle, Sentinel is in position P1, with the search volume shown in red. As it progresses through the cycle, Sentinel scans its $FOR$, exiting the cycle at position P2, with search volume shown in blue. The purple region is the overlap. The ratio of the fraction of NEO in the blue area to that in the purple is the change in asteroid fraction that occurs during one cadence cycle (right)—45% for Sentinel for asteroids of 140 m diameter. Variations from High to Low albedo scenarios for LSST and Cubesat-5 are indicated by shaded regions; there is no difference among scenarios for either Sentinel or NEOCam.

5.1.2.   Noise and Background Confusion

All telescope systems must recognize asteroids from digital images, assemble the images into tracklets, and then derive orbital elements. Noisy sensor data could create false positives. Internal reflections in the telescope and other factors can confound detection.

Background confusion will limit some observations. In the case of Pan-STARRS1, residual charge in the sensor CCDs from bright stars, a form of time-delayed background confusion, posed a problem (Denneau et al., 2013). These effects can be modelled, but I did not do so for this study.

### 5.1.3. Trailing Losses

Asteroid movement can cause trailing losses, which occur when the captured photons from the asteroid are distributed across multiple pixels during an exposure. This effect potentially lowers the *SNR* below detection thresholds. Calculation of the extent of trailing losses depends on per-orbit simulation and thus is the beyond the scope of this analysis. However, on a qualitative basis trailing losses should have at most modest effects on the performance of these telescopes. The Cubesat-5 constellation uses synthetic tracking, so by definition it experiences no trailing losses.

LSST's very short exposure time (15 s) and very wide field of view mitigate trailing losses. In addition, LSST will use a deep search region that ensures that many of the asteroids it detects will be at $r_{as} \geq 1.5$ au (see Fig. 15) and thus will have lower apparent motion than those at smaller $r$, although the strength of this effect depends on the viewing geometry.

Telescopes can look to the zenith to observe asteroids at opposition, during which their apparent motion slows, then retrogrades. This approach offers additional opportunities, assuming that the opposition is within the search volume. NEOCam currently plans to exclude the zenith from its FOR, presumably to optimize cadence.

Trailing losses will almost certainly be important for some subset of the asteroid population, but the magnitude of that subset is a topic that only a detailed, orbit-by-orbit simulation can address.

### 5.1.4. Weather

LSST is a terrestrial telescope, and weather and cloud cover could hamper its observations and have a detrimental effect on detection rate. Weather effects have been modeled previously (Ivezić et al., 2006), and a future, more detailed model could include them.

### 5.1.5. Zodiacal Dust

The IR telescopes are affected by background noise from zodiacal dust, which is anisotropic. The assumptions used in this study may overestimate or underestimate the noise. A more complete analysis would incorporate a full zodiacal-dust model.

### *5.2. Comparison to Previous Results*

Detailed simulation studies of each telescope have been published (Ivezić et al., 2006; Lu et al., 2013; Mainzer et al., 2015; Shao et al., 2015), but because they rely on code that is unavailable it is hard to determine the degree to which they simulate all factors to the same level of detail. As shown here, even some of the most basic assumptions—such as the thermal model for IR telescopes and the range of solar elongation that are observable from Earth—have been subject to disparate assumptions and treatment.

Because the albedo distributions of small NEOs are not well constrained by direct observation, it is also possible that the assumptions used for modeling the performance of telescopes are not compatible. They may effectively assume different NEO populations with respect to the distributions of key parameters, such as $G$, $p_v$, and $p_{ir}$. Consistent use of such details is crucial when comparing the likely performance of alternative systems.

Many prior studies focused on the GEB. The top-line result from these studies is that each of the systems can perform the GEB within a certain time frame. Unfortunately, GEB completion time is a function of both orbital dynamics and the telescope, so comparing completion dates alone is not that informative. Moreover, the GEB is ambiguous because the 90% criterion does not specify the nature of what is in the 10% that may be omitted. Depending on the nature of the telescope, the missed asteroids might be the darkest ones (*i.e.*, those having low $p_v$, $G$), those that have unfavorable orbits, or some combination.

Prior studies each conclude that the proposed telescope system will make significant progress, or complete the GEB within their mission timeline (Ivezić et al., 2006; Lu et al., 2013; Mainzer et al., 2015; Shao et al., 2015). The analysis methods used here cannot either refute or confirm the results. Many core assumptions (such as $G$, $p_v$, and $p_{ir}$) for these studies are not published, and those that are appear to differ from each other.

From my analysis of search volumes, for example, it seems likely that Cubesat-5 will perform poorly on dark asteroids. Other factors seem unlikely to remove the advantage in search volume that the other systems have over Cubesat-5. As a result, it seems plausible that when Cubesat-5 completes the GEB, the 10% of asteroids it misses will likely have different characteristics than those missed by one of the IR telescopes that can see low $p_v$ asteroids better than bright ones.

Despite these differences, the results of this study are not inconsistent with previously published results. Although NEOCam emerges with the smallest overall number of asteroids in its search volume, that number is likely quite consistent with it being able to complete the GEB.

Sentinel and NEOCam were designed for the purpose of completing the GEB, and as a result they are likely to have been made just good enough (from a cost perspective) to accomplish the mission. LSST was designed for more demanding observations of deep space objects, so its search is much deeper than would be needed for an NEO survey alone. This design advantage explains why the fraction of asteroids in its search volume is an order of magnitude higher than that of the other telescopes.

The Cubesat-5 constellation pioneers a new and potentially disruptive approach to asteroid observation by using multiple small satellites, each having a very small (10 cm) aperture telescope. The fivefold increase in breadth that comes from the constellation gives it exceptional searching ability for such a small aperture. But there is an apparent trade-off in its ability to observe small or dark asteroids.

## 5.3. *Tumbling Asteroids*

A surprising finding of this study is that asteroids that are tumbling quickly or have high thermal inertia, or both, may be very challenging for an IR telescope to detect, particularly if they are small. It is well known that some asteroids have high thermal inertia, likely due to a bare rock surface. It is also known that some asteroids rotate rapidly, and that is why the FRM model has been in use for more than two decades. It logically follows that a quickly tumbling asteroid having high thermal inertia can be modeled by the Tumble model presented in section 2.3. (To handle asteroids that rotate quickly about one axis and tumble slowly around another axis, one could interpolate between Tumble and the GFRM models.) It is unknown what fraction of NEOs, particularly at small diameter, may have these properties.

If even a small percentage do, that fact greatly complicates reaching a high-completion threshold result, such as the GEB. For example, if 10% of all NEOs of diameter 140 m of greater meet the criteria for Tumble, then it would be virtually impossible for an IR telescope to complete the GEB. Previous simulations for completion already assumed that 10% of target asteroids would be missed for other reasons. If an additional 10% are very hard to detect, one would expect the maximum completion percentage to be closer to 80%.

A smaller percentage (*e.g.*, 1% to 2%) of fast-tumbling/high thermal inertia asteroids could still put a burden on the search and would almost certainly extend the time to completion. Because a $d = 478$ m Tumble asteroid has the same sensor response (for the IR telescopes studied here) as a $d = 140$ m asteroid with the same physical properties, the asteroids missed by an IR telescope could be large—perhaps as large as 450 m in diameter. Fortunately, tumbling asteroids are readily detected by visual-band telescopes. This suggests that any future strategy for NEO searching should involve both visible and IR, because IR alone has this vulnerability.

It is possible that IR observations of such asteroids would misestimate their diameter. Visual albedos are typically derived from IR-determined diameters, so there is an intriguing possibility that at least some fraction of the asteroids currently identified as high $p_v$ objects are actually larger than realized and have lower true values of $p_v$. This is a topic of ongoing research.

## 6.     Conclusions

Although the simple calculations done here have utility, the best next step would be to create an open and accessible simulation platform for comparing the different systems in detail. Of the telescope projects reviewed here, only LSST has secured funding and begun construction. A detailed simulation framework could act as a catalyst to help more projects move forward toward completion.

A simulation platform could also help answer the important question of what the goal of NEO search should be. The original GEB now seems unlikely to be completed by the original 2020 deadline, but it is already clear that something more ambitious is needed. The start of construction on LSST marks an interesting turning point because it makes it virtually certain that the GEB will be completed in the foreseeable future. Even those most skeptical of LSST as a NEO search tool should take it into account.

The other proposed systems have yet to adapt to the reality that LSST is proceeding. A recent paper on NEO alternatives (Shao et al., 2015) does not even mention LSST, despite discussing other proposed systems in detail and quoting extensively from the NRC study that found LSST to be the most cost-effective option. While some skepticism is warranted with any ambitious and as yet unproven project, many of the scientific points that critics of LSST use to buttress their argument are less clear-cut upon close examination than they first seem.

A space-based mission could proceed regardless of LSST's search for NEOs, with the idea that it could help find asteroids more quickly than LSST will do alone. Such a proposal would need to be supported with careful study to assess the actual value added, however. Racing LSST to find the same asteroids a few years before it does would not have much value in planetary defense. This is particularly true given the front-loaded detection rate that LSST enjoys; it may complete one-third of its search within the first week of operation.

A better strategy might be to choose a mission that compliments LSST by exceeding its capabilities for a class of asteroids. Quantitative study by an open simulation framework could further characterize the marginal advantage of deploying a second space based system. It may also point the way to different design tradeoffs to optimize a future NEO search mission that works in partnership with LSST.

## Acknowledgements

I want to acknowledge extremely open and helpful discussions from Steve Chesley, Zeljko Ivezic, Mario Juric, Lynne Jones, Jordin Kare, Amy Mainzer, Harold Reitsema, Tom Statler, Michael Shao, Tony Tyson, Alan W. Harris, Alan W. Harris, Bruce Hapke, and Peter Veres and an anonymous reviewer.

No outside funding was received for this study.